\documentclass[fleqn,usenatbib,useAMS]{mnras}

\usepackage{newtxtext,newtxmath}
\usepackage[T1]{fontenc}

\DeclareRobustCommand{\VAN}[3]{#2}
\let\VANthebibliography\thebibliography
\def\thebibliography{\DeclareRobustCommand{\VAN}[3]{##3}\VANthebibliography}

\usepackage{graphicx}	% Including figure files
\usepackage{amsmath}	% Advanced maths commands
\usepackage{lscape} 
\usepackage{multirow}
\usepackage{rotating}
\usepackage{footmisc}
\usepackage{soul}
\usepackage[normalem]{ulem}
\providecommand{\logg}{cm\,s$^{-2}$}
\newcommand{\kms}{\,km\,s$^{-1}$}
\newcommand{\mos}{\,m\,s$^{-1}$}
\newcommand{\rev}\textbf{}
\newcommand{\radd}{\,rad\,d$^{-1}$}
\title[Planets around young Solar-type stars.]{Planets around young active Solar-type stars:\\ Assessing detection capabilities from a non stabilised spectrograph.}

\author[A. Heitzmann et al.]{
A. Heitzmann,$^{1}$\thanks{E-mail: alexis.heitzmann@usq.edu.au}
S.C. Marsden,$^{1}$
P. Petit,$^{2}$
M. W. Mengel,$^{1}$
D. Wright,$^{1}$
M. Clerte,$^{1}$
\newauthor  I. Millburn,$^{1}$
C.P. Folsom,$^{3,4}$
B. C. Addison,$^{1}$
R. A. Wittenmyer,$^{1}$
I.A. Waite$^{1}$
\\
$^{1}$Centre for Astrophysics, University of Southern Queensland, West Street, Toowoomba, QLD 4350 Australia\\
$^{2}$Institut de Recherche en Astrophysique et Planétologie, Université de Toulouse, CNRS, CNES, 14 avenue Edouard Belin, 31400 Toulouse, France\\
$^{3}$Department of Physics \& Space Science, Royal Military College of Canada, PO Box 17000 Station Forces, Kingston, ON, Canada K7K 0C6\\
$^{4}$Tartu Observatory, University of Tartu, Observatooriumi 1, Tõravere, 61602 Tartumaa, Estonia
}

\date{Accepted XXX. Received YYY; in original form ZZZ}

\pubyear{2020}

\begin{document}
\label{firstpage}
\pagerange{\pageref{firstpage}--\pageref{lastpage}}
\maketitle

\begin{abstract}
Short-orbit gas giant planet formation/evolution mechanisms are still not well understood. One promising pathway to discriminate between mechanisms is to constrain the occurrence rate of these peculiar exoplanets at the earliest stage of the system's life. However, a major limitation when studying newly born stars is stellar activity. This cocktail of phenomena triggered by fast rotation, strong magnetic fields and complex internal dynamics, especially present in very young stars, compromises our ability to detect exoplanets. In this paper, we investigated the limitations of such detections in the context of already acquired data solely using radial velocity data acquired with a non-stabilised spectrograph. We employed two strategies: Doppler Imaging and Gaussian Processes and could confidently detect Hot Jupiters with semi-amplitude of 100\,\mos buried in the stellar activity. We also showed the advantages of the Gaussian Process approach in this case. This study serves as a proof of concept to identify potential candidates for follow-up observations or even discover such planets in legacy datasets available to the community.
%{\expandafter\string\the\font}
\end{abstract}


%%%%%%%%%%%%%%%%%%%%%%%%%%%%%%%%%%%%%%%%%%%%%%%%%%

%%%%%%%%%%%%%%%%% BODY OF PAPER %%%%%%%%%%%%%%%%%%

\section{Introduction}

In the quest to understand the processes governing planetary system development, the peculiar case of short-orbit gas giants (i.e., Jovian and sub-Jovian exoplanets, orbiting their star with periods of less than a few weeks that we will refer as hot Jupiters or HJs hereafter), is a real challenge as classical theories describing their formation and evolution do not predict their presence in the vicinity of their parent star. Although representing a significant fraction of all exoplanets discovered (between 10 and 15 per cent \footnote{\url{https://exoplanetarchive.ipac.caltech.edu/}}), their true occurrence rate is estimated to be around 1 per cent for mature Solar-type stars \citep{Wright2012}. Even though this discrepancy can be explained through observing biases, their scarcity raises the question of the formation channel generating this population.\par 
In the most accepted explanation, future HJs form in the colder region of the protoplanetary disk (more than a few au) and later experience orbital decay to eventually reach a close-in orbit. Two migration mechanisms are proposed: gas disk migration  (see \citealt{Baruteau2014} for a review), where the planet migrates inward as the result of angular momentum exchange between the gas giant and the disk, and high eccentricity tidal migration. In this last scenario, the planet is sent to a highly eccentric orbit following a strong perturbation (planet-planet scattering, e.g. \citealt{Chatterjee2008} or secular interactions, see \citealt{Beauge2012}, \citealt{Petrovich2016}, \citealt{Petrovich2015} and \citealt{Hamers2017} for the different proposed mechanisms). Now being close enough to the star at periastron, tidal forces exerted by the star act to circularise the planet's orbit.\par 
Confronting migration theories is in-situ formation, where the HJ forms in the vicinity of the host star and remains in close-orbit. This explanation has been historically rejected as it sets restrictive constraints on the inner stellar disk, i.e. there must be enough available material to form the core of these gas giants and that core forming process needs to be completed before the star depletes all the gas from the area for the future HJ to successfully accrete its gaseous envelope. Due to these constraints, it is unlikely to occur according to the \emph{Solar nebula theory}, assuming a disk composition similar to the one that gave birth to our Solar system \citep{Perryman2011}. Now realising that our Solar system may be far from being the norm in the great diversity of planetary systems, in-situ formation has come back under the spotlight \citep{Batygin2015,Boley2015}. Recent studies, such as \cite{Bailey2018} or \cite{Dawson2018}, suggest that HJs could have a different origin in different systems and/or that a combination of the proposed mechanisms could be at play. \par
In their review paper, \cite{Dawson2018} propose to test the different theories by searching for correlations between properties of HJs and their parent stars. Among the 15 studied properties, two are flagged as requiring further observations: HJ obliquities and host star ages. This paper focuses on the latter.\par
Studying young stars is a privileged approach as it would help to discriminate between early-stage mechanisms such as in-situ formation or gas disk migration versus more prolonged and late-stage mechanisms, like high eccentricity migration. However, \cite{Dawson2018} warn that high-eccentricity migration driven HJs could arrive in close-in orbit fairly early in the system's formation, showing that the dependence of these mechanisms with stellar age is not yet completely clear. Therefore, young stars (< 20-50 Myr), and even more so, younger (< 10 Myr) low mass (< 3 $M_{\odot}$) T Tauri pre main sequence (PMS) stars as defined in \citet{1989A&ARv...1..291A} are probably the best candidates. \par 
Unfortunately, with the exception of direct imaging surveys, the youngest stars are typically systematically avoided when searching for exoplanets, as they exhibit particularly strong intrinsic variability, or stellar activity. For such stars, this activity-induced correlated noise results primarily from surface brightness features, linked to complex internal processes and a strong magnetic field. Surface features yield spurious radial velocity (RV) signatures that generally completely mask exoplanet signatures, hence preventing their discovery. Additionally, \citealt{Nava2019} showed that activity can generate unexpected spurious peaks in a periodogram analysis, which could lead to false positives if no adequate treatment of the activity is applied. \par
Filtering, or mitigating this stellar activity becomes crucial if one hopes to find traces of exoplanets orbiting young active stars. It is also important to note that effective activity mitigating strategies are key in the search of Earth sized planet around less active stars. In those cases, both the activity level and the planetary signature are up to 2 orders of magnitude smaller, but present a similar situation in relative terms. However, it is still slightly different as additional phenomena are also at play (i.e. granulation, pulsations). The exoplanet community is actively trying to develop and assess these strategies (see \citealt{cabot20}).\par
Available data on planets with periods less than 15\,days orbiting very young stars (< 50 Myr) is very scarce. Six planet-hosting stars have been found from transits \citep{David2016,David2019,Newton2019,Rizzuto_2020,Plavchan2020,bouma20} and three from RV searches: CI Tau b \citep{Johns-Krull2016,Flagg2019}, V830 Tau b \citep{Donati2015,Donati2016,Donati2017} and TAP 26 b \citep{Yu2017b}. Recently, however, the existence of both V830 Tau b and CI Tau b have been challenged by \citealt{Damasso2020} and \citealt{Donati2020}. V830 Tau b and TAP 26 b were found by the MaTYSSE (Magnetic Topologies of Young Stars and the Survival of massive close-in Exoplanets) observation programme in a sample of 33 weak-line T Tauri stars \citep{Yu2017a}. If real, these 2 planets would indicate a fraction of HJs as high as 6 per cent for newly born stars. In this context, it is crucial to carry on the search for close-in gas giants around young stars to better estimate their occurrence rate at that stage. \par
In this paper, we investigated the case of searches for short-period gas giants orbiting very young and active stars solely using RV data. More specifically, we injected various RV signatures mimicking single circular planet systems behind real data of the young active G dwarf HD 141943 (not known to host a massive planetary companion) and assessed our detection limits using two distinct strategies: Doppler Imaging (DI) activity filtering (Section~\ref{subsec:Method1}) and Gaussian Process (GP) Regression (Section~\ref{subsec:method_GP}). \par
Although already used in the past (DI + GP in \citealt{Donati2016,Donati2017,Yu2017b,Yu2019,Klein2020} and GP in most exoplanet searches for the past few years), testing the respective performance of these two methods in legacy datasets has not been performed. We note that the underlying data were not optimised to search for exoplanets and was obtained using a non stabilised spectrograph (e.g. with $\approx$50-100\,\mos uncertainty on radial velocities). The limitations we describe should therefore be significantly improved with RV stabilised datasets. However, they provide a strong baseline for what is achievable and are useful to investigate other datasets of this nature already available (i.e. in the Bcool \citep{Marsden2014} or TOUPIES \citep{Folsom2016,Folsom2018} surveys). We compared our results to the planet 'hide and seek' study done on the same star with no specific treatment for stellar activity \citep{Jeffers2014}.\par
This paper is organised as follows. Details on the techniques used to reduce the data, more specifically to get from raw spectra to radial velocities are given in Section~\ref{sec:DataAnlaysis}. We then cover the methods addressing stellar variability in Section~\ref{sec:filtering}. Part~\ref{sec:maindataset} of this paper focuses on our reanalysis of HD 141943's raw dataset. Section~\ref{sec:simulations} explains how we set up our simulated datasets, and results from the analysis are laid out in ~\ref{sec:Results}. Finally, we give our conclusions and future prospects in Section~\ref{sec:Conclusions} and ~\ref{sec:futurework}.

\begin{figure}
	\includegraphics[width=\columnwidth]{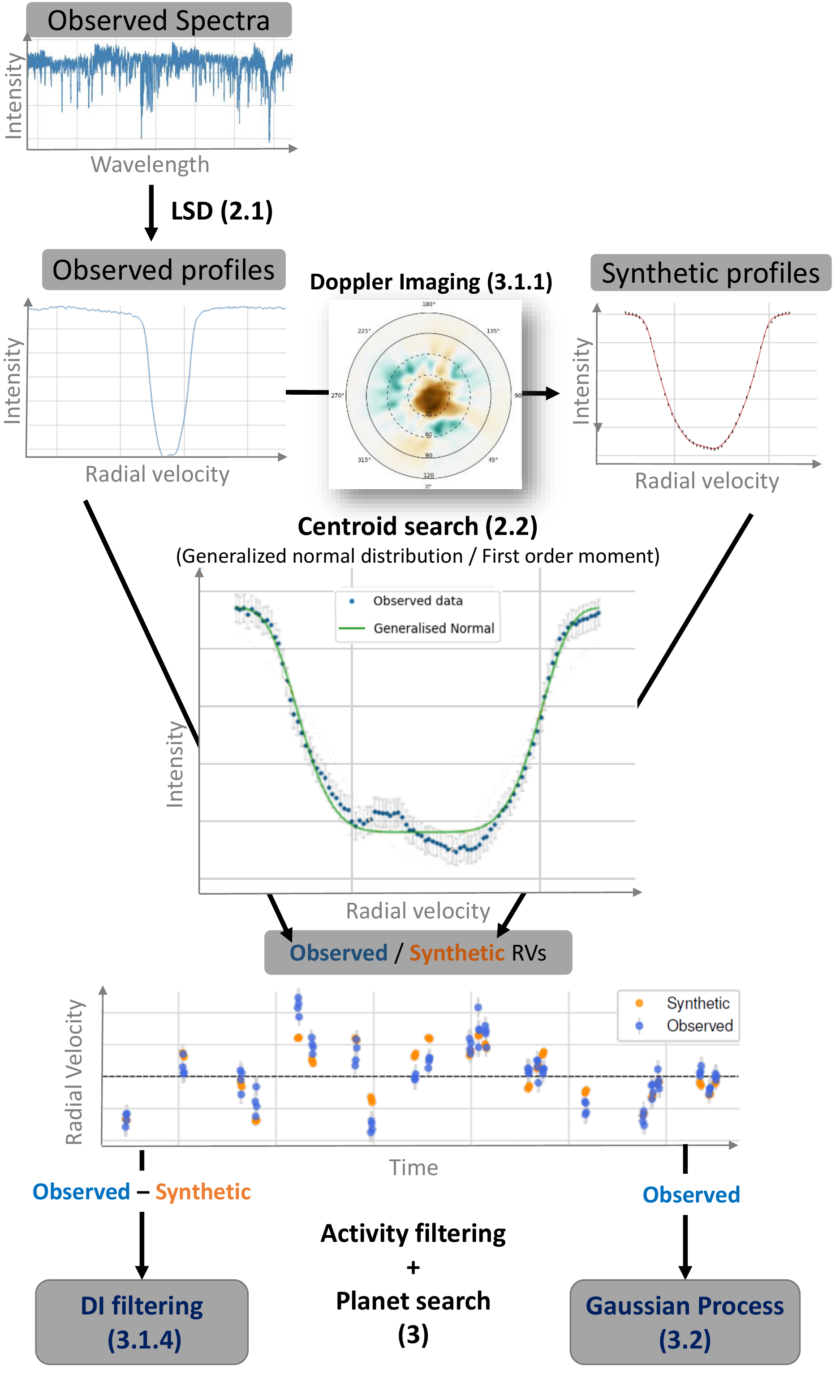}
    \caption{Diagram of the data analysis procedure, from raw spectra to periodic signature identification. Each block is a step of the process. Bold text and the associated numbers in parenthesis respectively indicate the method used to progress from one block to the next and the section of this paper detailing the corresponding process. Each point on the bottom plot results from the analysis of a single spectrum using the entire procedure described here. }
    \label{fig:diagram}
\end{figure}

\section{Data analysis}
\label{sec:DataAnlaysis}

\subsection{From spectra to line profiles}
\label{sec:spectratoline}
Both methods we utilised to disentangle stellar activity from planetary signals take radial velocity time series as input. The extraction of RV values from raw stellar spectra were performed by finding the centroid (described in Section~\ref{subsec:RV}) of a `mean line profile' obtained using Least-square Deconvolution (LSD, \citealp{Donati1997}, \citealp{Kochukhov2010}).
%% Not done in the end
% {\color{red} or a cross-correlation function (CCF) for the Gaussian Process (GP) Regression. The choice of a CCF method over the LSD is to ensure the accuracy of error bars as these are always overestimated when performing LSD. However, our first method uses LSD as it is a requirement for Doppler Imaging}. \par
LSD convolves an observed stellar spectrum with a spectral line mask. Given an appropriate mask, the result is an enhanced peak signal to noise ratio (S/N) `mean line profile' exhibiting stellar activity induced line features. We chose the stellar mask best matching our star in the list of masks designed in the scope of the Bcool survey \citep{Marsden2014} using VALD \footnote{\url{http://vald.astro.uu.se/}} for a star with an effective temperature of $T_{\mathrm{eff}}$ = 6000 K, surface gravity of log~$g$ = 4.5\,\logg\, and [$\mathrm{Fe/H}$] = +0.2. Only spectral lines deeper than 20 per cent of the maximum line depth were kept for the LSD computation, yielding a total of 4097 lines. The outcome was a S/N increase from $\approx$ 50-100 for the observed spectra (depending on the spectrum and spectral order considered) to $\approx$ 1000 for the LSD mean line profiles.

\subsection{From line profiles to radial velocities}
\label{subsec:RV}
Classically, each RV is taken to be the mean of a Gaussian profile fitted to the obtained line profile. However, for active stars, the distortion and here the 'flat-bottom' (see the centre plot on Figure~\ref{fig:diagram}) of the line shows that a Gaussian fit is not suitable. We considered two alternatives. \par
Firstly, we chose a Generalised Normal Distribution (GND, \citealt{Nadarajah2005}), as shown in green on the central plot of Figure~\ref{fig:diagram} and described by the following p.d.f:

\begin{equation}
    \mathrm{GND}(x) = \frac{\beta}{2\sigma\Gamma\left(\frac{1}{\beta}\right)}\exp\left(-\left|\frac{x-\mu}{\sigma}\right|^{\beta}\right)
\end{equation}

where $\Gamma$ denotes the gamma function, $\mu$ the position parameter (mean), $\sigma$ the scale parameter, and $\beta$ the shape parameter. $\beta < 2$ results in wings more extended than a normal distribution and a sharper distribution peak. When $\beta = 2$, the GND becomes a Gaussian distribution (where $\sigma$ is the standard deviation). For $\beta > 2$, the distribution yields wings less extended than a normal distribution and tends to a uniform distribution as $\beta\to\infty$ . This grants more flexibility to the distribution resulting in a better fit to broadened profiles. Error bars on the GND parameters are given by the fitting method. The centroid $\mu$ and the associated error bars for each LSD profile constituted our RV time series. \par
Secondly, we derived RV values using the first-order moment (generalised centroid, FOM hereafter) of each LSD profile, computed as:
\begin{equation}
    \mathrm{RV} = \frac{\int\left( I_c-I(\nu)\right)\nu d\nu}{\int\left( I_c-I(\nu)\right)d\nu}
\end{equation}
with $I(\nu)$ the intensity of the profile at radial velocity $\nu$ and $I_c$ the continuum level. Here, error bars were propagated using the LSD derived uncertainties. We note that results given by FOM are sensitive to integration limits (i.e. the limits on the line profiles used to compute it). This is further described in Section~\ref{subsec:RV_extractin}.

\section{Stellar activity filtering}
\label{sec:filtering}
Stellar activity distorts line profiles, causing a shift in the line's centroid and therefore in the measured RV. Modelling the activity thus aims to correct for these distortion induced shifts.

\subsection{Method \#1: Filtering activity using Doppler Imaging}
\label{subsec:Method1}
Section~\ref{subsec:DopperImaging} describes Doppler Imaging, representing the core of our filtering process, following \cite{Donati2014}.  Sections~\ref{subsec:ZDI} and ~\ref{subsec:diffrot} describe magnetic imaging (ZDI) and differential rotation, complementary to the DI technique. The actual filtering process is described in~\ref{subsec:DI filtering}.

\subsubsection{Doppler Imaging}
\label{subsec:DopperImaging}
Doppler Imaging (DI) is a tomographic technique that, for rapidly rotating stars ($v\sin i \gtrsim 10$\,\kms), uses spectroscopic observations to infer the brightness features at their surface \citep{Brown1991,Donati1997_2}. 
Practically, a time-series of observed pseudo-line profiles obtained through LSD are iteratively adjusted using a tomographic algorithm. Irregularities in the profiles are interpreted as surface bright/dark spots that enhance/block Doppler shifted light due to stellar rotation. Then, iteratively, synthetic profiles, derived from the DI surface maps, are fit to the observed ones. To reach a unique solution to the ill-posed problem of DI inversion (as a single line profile can be generated from different surface map solutions), a maximum entropy selection of the solution is adopted (i.e. minimising the information content of the brightness map), while ensuring that the $\chi^2$ is kept below a defined threshold. This is done following the routine of \cite{Skilling1984} and using the entropy as defined in \cite{Hobson1998}. Further details can be found in Appendix B of \cite{Folsom2016}. The model output is constituted of a synthetic set of LSD profiles, and of the brightness surface map producing this spectral information.\par 
Synthetic line profiles are obtained by integrating the Doppler-shifted flux (due to the rotation of the star) emerging from each point of the visible hemisphere. This flux is scaled according to the local surface cell projected area, brightness and limb darkening. The local line profiles are calculated using a Voigt profile, a convolution of a Gaussian and a Lorentzian profile. \par
Output products of DI include a set of synthetic profiles and a surface brightness map (or a magnetic map for Zeeman Doppler Imaging, see next section). The use of DI also enables us to constrain the stellar fundamental parameters by selecting the parameter values that optimise the brightness model (i.e., inclination of the stellar rotational axis with respect to the line-of-sight $i$, line-of-sight projected equatorial rotation velocity $v\sin i$, stellar equatorial rotation period $P_{\mathrm{eq}}$, stellar mean radial velocity $\overline{RV}$ and differential rotation $\mathrm{d}\Omega$) and line profile parameters (i.e., line depth, Gaussian and Lorentzian equivalent widths). The DI analysis of HD 141943 is described in Section~\ref{subsec:refinedstellarparam} and Figure~\ref{fig:maps_comparison}.

\subsubsection{Zeeman Doppler Imaging}
\label{subsec:ZDI}

Although Zeeman Doppler Imaging is not part of the filtering process, it is similar to the stellar mapping process and is therefore described here. \par
Similarly to DI, Zeeman Doppler Imaging (ZDI, e.g. \citealt{Semel1989}) is a technique that uses polarimetric information (i.e. Stokes V LSD profiles) to reconstruct the magnetic field structure at the surface of the star. Here we used a spherical harmonic expansion to describe the large-scale components of the magnetic field (i.e., poloidal and toroidal, \citealt{Donati2006}). The Zeeman effect allows one to infer the strength and direction of the surface magnetic field, provided one has high enough S/N line profiles, rendered possible by the LSD technique. Like DI, solving for a magnetic field configuration is an ill-posed problem and ZDI also relies on maximum entropy image reconstruction. The ZDI analysis of HD 141943 is described in Section~\ref{subsec:refinedstellarparam} and Figure~\ref{fig:mag_map}.

\begin{figure}
	\includegraphics[width=\columnwidth]{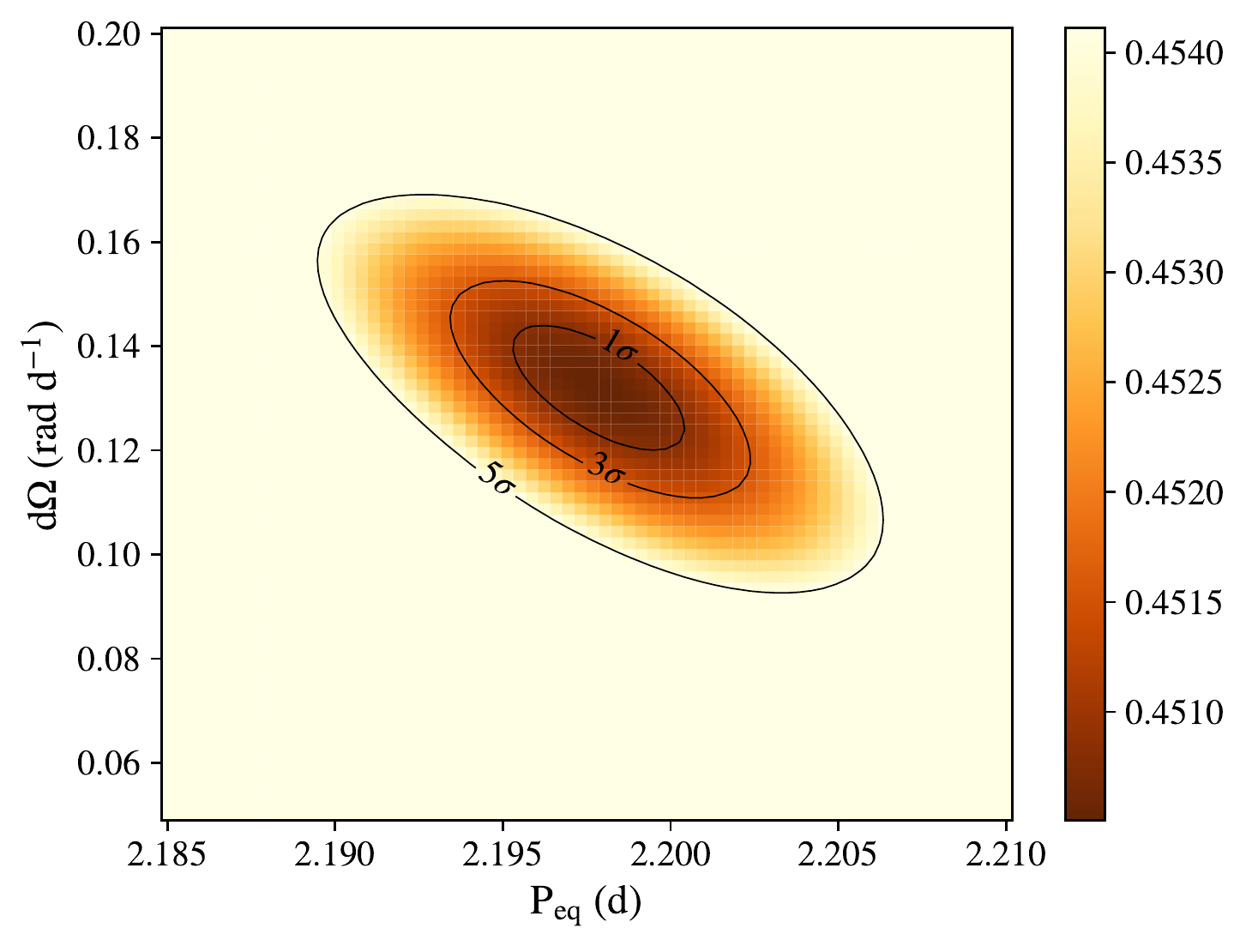}
    \caption{Reduced $\chi^2$ surface of the differential rotation (\radd) (y-axis) versus equatorial rotation period (x-axis), equivalent to the rotation frequency $\Omega_\mathrm{{eq}} = 2\pi / P_{\mathrm{eq}}$. Contours show confidence levels at 1, 3 and 5$\sigma$. Colour bar shows the reduced $\chi^2$ values.}
    \label{fig:Chisquaremap}
\end{figure}

\subsubsection{Surface differential rotation}
\label{subsec:diffrot}

The information used to generate a snapshot of the stellar surface through DI and ZDI often spans multiple stellar rotation cycles. Thus, the effect of differential rotation needs to be accounted for. The code we used models that differential rotation as a simplified solar-like differential rotation law:

\begin{equation}
    \Omega(\theta) = \Omega_{\mathrm{eq}} - \mathrm{d}\Omega \sin^2 \theta
\end{equation}
with $\Omega(\theta)$ the rotation rate at latitude $\theta$, $\Omega_{\mathrm{eq}}$ (=$\frac{2\pi}{P_{\mathrm{eq}}}$) the rotation rate at the equator and $\mathrm{d}\Omega$ the difference in rotation rate between the equator and the poles (i.e., the differential rotation). Following \cite{Petit2002} and \cite{Donati2003}, we explored the $\mathrm{d}\Omega$ and $\Omega_{\mathrm{eq}}$ parameter space, by running DI inversions for various values of the two parameters, looking for the doublet that optimizes the DI model, i.e. the $\mathrm{d}\Omega$ and $\Omega_{\mathrm{eq}}$ values that minimize the $\chi^2$ of our model at fixed entropy level. The resulting $\chi^2$ surface is used to derive our uncertainty on these two parameters. \par 
We performed our DI, ZDI and differential rotation analyses using the python \textsc{zdipy} code (see Appendix from \citet{Folsom2018} for a more detailed description of the code). The code has been adapted to run on Fawkes, the High Performance Computing (HPC) facility at the University of Southern Queensland. The HPC allowed us to quickly explore our parameter space. Practically, we varied the stellar parameters (up to 3 at a time) to find the best solution. By best solution we mean the set of parameters that fit our line profiles down to the target $\chi^2$ (< 1 due to the LSD process, see \citealt{Cang2020} for a similar case and follow up explanations) and also maximises the entropy value. The differential rotation analysis of HD 141943 is described in Section~\ref{subsec:refinedstellarparam} and Figure~\ref{fig:Chisquaremap}.

\subsubsection{Filtering the activity}
\label{subsec:DI filtering}
Following \cite{Donati2014}, we removed the stellar activity contribution by subtracting the RV time series derived from our modelling of the activity alone (i.e., the synthetic profiles centroids obtained from DI) from the values obtained from the raw/observed LSD profiles (i.e. the observed/raw LSD profiles centroids). We assumed that for very active stars, stellar variability is in a first approximation entirely due to features present on the stellar surface. We then searched for periodicity in the resulting filtered RVs, utilising a Lomb Scargle (LS) periodogram (\citealt{Lomb1976,Scargle1982}). We point out that residuals exhibit some red noise leftovers, while LS periodograms are designed for uncorrelated/white noise \citep{Vanderplas2018}. We keep this approach here for maximal consistency with previous papers of \cite{Donati2014,Donati2016,Yu2017b,Yu2019}.
\par
The nature of stellar variability (i.e. correlated/red noise) combined with the imperfect filtering (see Figure~\ref{fig:per_5}) of the activity using DI, results in residuals exhibiting some red noise leftovers. As LS periodograms are designed for uncorrelated/white noise \citep{Vanderplas2018}, this approach is limited and should not be used alone to claim a planet detection. To assess significance of a detection, we use the false alarm probability (FAP)\footnote{The FAP limit indicates the likelihood that a peak caused by random fluctuations in the data would reach a given height/power (see dashed lines in Figures~\ref{fig:per_5}, \ref{fig:periodograms32_36} and \ref{fig:per_22} ). However, it does not indicate the probability of a dataset to have a periodic component given the data.}. To compute the FAP levels, we used the Baluev approximation (see \citealt{Baluev2008}). We also tried a bootstrap approach, which yielded very similar results.\par

\begin{table}
	\centering
	\caption{Prior distributions of parameters used for the Gaussian Process regression. The right column gives describes the prior for each parameter of the model. J(\emph{min},\emph{max}) stands for Jeffreys priors, MJ(\emph{max},\emph{knee}) for Modified Jefferys priors, $\mathcal{N}$(\emph{mean},\emph{std}) for Gaussian priors and $\mathcal{U}$[\emph{min},\emph{max}] for Uniform priors. $\overline{\sigma}_{\mathrm{RV}}$ is the mean of the RV uncertainties. $RV_{\mathrm{max}}$ is the largest absolute value in the dataset and $RV_{\mathrm{std}}$ the standard deviation of all RV values.}
	\label{tab:GPpriors}
    \begin{tabular}{ll}
    \hline
    Parameters & Priors  \\
    \hline
    \hline
    Stellar activity & \\
    $\theta_1$ (\kms)& MJ(1.5$\times\,RV_{\mathrm{max}}$,$\overline{\sigma}_{\mathrm{RV}}$)\\
    $\theta_2$ (d)& J(1,100)\\
    $\theta_3$ (d)& $\mathcal{N}$(2.2,0.05) \\
    $\theta_4$ [0:1]& $\mathcal{U}$[0:1] \\
    \hline
    Planet & \\
    $K$ (\kms) & MJ(2$\times\,RV_{\mathrm{max}}$,$\overline{\sigma}_{\mathrm{RV}}$)\\
    $P_{\mathrm{orb}}$ (d) & J(0.1,15) \\
    $\Phi$ [0:1] & $\mathcal{U}$[0:1] \\
    \hline
    Telescope and Noise & \\
    $RV_\mathrm{0}$ (\kms)& $\mathcal{U}$[-$RV_{\mathrm{max}}$:$RV_{\mathrm{max}}$]  \\
    $\mathrm{\sigma_s}$ (\kms)& MJ($RV_{\mathrm{std}}$,$\overline{\sigma}_{\mathrm{RV}}$)  \\
    % \hline
    % \multicolumn{2}{l}{\textsuperscript{a}\footnotesize{\cite{Marsden2011}}}\\
    % \multicolumn{2}{l}{\textsuperscript{b}\footnotesize{\cite{Marsden2011}, agreeing with the results from this study}}\\
    % \multicolumn{2}{l}{\textsuperscript{c}\footnotesize{This study}}
    \end{tabular}
\end{table}

\subsection{Method \#2: Modelling the activity using  a Gaussian Process regression}
\label{subsec:method_GP}
Our second approach uses a Gaussian Process (GP hereafter) Regression to model the stellar activity induced RV and its temporal evolution as first suggested in \cite{haywood_planets_2014} and \cite{rajpaul_gaussian_2015}. The GP regression treats stellar activity as Gaussian red (correlated) noise. This Bayesian approach is driven by the data points, considered to be random correlated Gaussian variables and the covariance matrix \textbf{C}, specifying the correlation between each pair of data points. Following \cite{haywood_planets_2014} we computed each entry $\mathrm{\boldsymbol{\mathrm{C}}_{ij}}$ of this co-variance matrix using the following physically driven quasi-periodic kernel made of a sinusoidal component to account for the rotation of the star combined with an exponential component for the surface features appearance/decay:

\begin{equation}
\label{eq:kernel}
\mathrm{\boldsymbol{\mathrm{C}}_{ij}} = \theta_1^2 . \exp \left[ - \frac{\left( t_i - t_j\right)^2 }{\theta_2^2}- \frac{ \sin^2\left( \frac{\pi (t_i-t_j)}{\theta_3}\right)}{\theta_4^2} \right] + \left( \sigma_i^2 + \sigma_s^2 \right)\delta_{ij}
\end{equation}
Where the four hyper-parameters can be interpreted as:
 \begin{itemize}
     \item $\theta_1$ (\kms) : Semi-amplitude of the activity RV signature.
     \item $\theta_2$ (d) : Decay parameter, or typical surface feature lifetime.
     \item $\theta_3$ (d) : Recurrence timescale, expected to be very close to $P_{\mathrm{eq}}$.
     \item $\theta_4$ [0:1] : Smoothing parameter or amount of high frequency in the signal. Smaller and larger value of $\theta_4$ respectively indicates variations on a longer and shorter timescale. From experience (\citealp{Haywood2018,Jeffers2009}), light curves and RV timeseries exhibit values of around 0.3 to 0.4 for this parameter. We chose a uniform prior that guarantees to largely encompass these values.
 \end{itemize}
$\sigma_i$ is the uncertainty of datapoint $i$ and $\sigma_s$ an extra white, uncorrelated noise parameter accounting for variations due to other sources and not explicitly captured by the model. $\sigma_i$ and $\sigma_s$ were added in quadrature and only applied to the diagonal of our matrix (i.e. variance of the datapoints). \par
Our global model is the sum of: the GP model accounting for the stellar activity ($RV_{\mathrm{GP}}$), a sinusoid for the circular planetary signature ($RV_{\mathrm{pla}}$) and a constant offset ($RV_\mathrm{0}$): \par
\begin{equation} 
    \begin{split}
        \mathrm{RV_{tot}} = &\quad \; RV_{0} \\
         & + \;RV_{ \mathrm{GP}}(t,\theta_1,\theta_2,\theta_3,\theta_4,\sigma_i,\sigma_s) \\ 
         & +\; \mathrm{RV}_{\mathrm{pla}}(t, K, P_{\mathrm{orb}},\Phi)
    \end{split}
\end{equation}
We ended up with a parameter space to explore containing 5 ($\theta_1$, $\theta_2$, $\theta_3$, $\theta_4$ and $\sigma_s$) + 3 $\times$ $n$ parameters + $\mathrm{RV}_\mathrm{0}$, for $n$ planets (i.e. 9 parameters for single planet model). Then, two aspects need to be considered in order to confidently claim the presence of a periodic signal in the data. The first part is parameter estimation, where we explore the parameter space yielding posterior distributions from which the most likely set of parameters, as well as their mean and uncertainty values, can be recovered. The second aspect is model selection, where we assess how much more likely a model containing one planet is, compared to one with stellar activity only. Commonly, parameter space exploration is performed using Monte Carlo approaches. Despite the efficiency of some algorithms (e.g emcee from \citealt{Foreman-Mackey2013}), the bottleneck of planet searches is usually model selection. \par 
Model selection is performed by comparing the \emph{marginal} likelihood (or evidence, $\mathcal{Z}$) of different models (i.e. activity only, activity with 1 planet, 2 planets, etc.). A detailed description of the evidence is given in Appendix~\ref{sec:appendixA}. Accurate estimation of this evidence is computationally expensive as it implies multi-dimensional integration over potentially large parameter spaces. Recently, \citealt{Nelson_2020} compared different methods for computing the evidence applied by different research groups. Although this was preliminary and would require follow up studies to completely generalise the results, some approaches proved to be more consistent than others.
Following their results, we developed our GP code using \textsc{pymultinest} \citep{Buchner2014}, a Python implementation of \textsc{multiNest} \citep{Feroz2009}. This Importance Nested Sampling algorithm estimates the evidence and provides, as a by-product, the posterior probabilities and can therefore also be used for parameter estimation. \par For the rest of this paper, when comparing models, we will refer to the Bayes Factor (BF) and/or the associated probability (p) in favour of a single planet model (model $\mathcal{M}_1$) over an activity only model (model $\mathcal{M}_0$):
\begin{equation}
\mathrm{BF} = \frac{\mathcal{Z}_1}{\mathcal{Z}_0} 
\label{eq:11}
\end{equation}
with $\mathcal{Z}_0$ and $\mathcal{Z}_1$ the marginal likelihood for $\mathcal{M}_0$ and $\mathcal{M}_1$. We used the metric of \cite{Jeffreys1961} (see Table~\ref{tab:Bayes factor}) to assess significance from the BF. \par
\subsubsection{Likelihood and Priors}
\label{subsec:priors}
Two ingredients are needed to recover the posterior probabilities; likelihoods and prior probabilities. \par
In our case, the natural logarithm of the likelihood (i.e. probability of the data given the model and its parameters, $\mathrm{p}(\boldsymbol{y}|\theta,\mathcal{M}_i)$ or $\mathcal{L}$), is given by: 
\begin{equation}
2 \ln \mathcal{L} = -n\ln(2\pi) - \ln \left( | \mathbf{C} | \right) - \mathbf{y}^T\left( \mathbf{C} \right)^{-1}\mathbf{y}
\end{equation}
With $\mathbf{y}$ the vector (of length $n$) containing the residuals after having removed both $RV_{\mathrm{pla}}$ and $RV_\mathrm{0}$ from the original RVs and $\mathbf{C}$ the co-variance matrix computed using our GP kernel from Equation~\ref{eq:kernel}. \par 
Our priors, physically motivated following \cite{Gregory2007}, are listed in Table~\ref{tab:GPpriors}. Because the evidence is dependent on prior probabilities, we emphasise the importance of favouring uninformative priors, such as uniform, Jeffrey's (uniform prior in logarithmic space, see \citealt{Gregory2007}) or Modified Jeffrey's (Jeffrey's prior, approaching a uniform distribution for values $\ll$ to the \emph{knee} parameter of the modified Jeffrey's prior to handle priors that have 0 as a lower boundary, also see \citealt{Gregory2007}), or at least priors independent of the studied data when previous and statistically valid knowledge is available in the literature. Using informative priors, without justification, would act to artificially boost the evidence. This is especially true for parameters that are not shared by the compared models (the planetary parameters in our case). The only informative prior we used here is $\theta_3$ as $P_{\mathrm{eq}}$ has been constrained from DI. \par
We ran \textsc{pymultinest} with an efficiency of 0.3 and 2000 live points (see \citealt{Nelson_2020}). For each run, the parameter search drew between $\approx$ 50,000 samples from the posterior for the model with no activity and $\approx$ 150,000 for the single-planet model. Details of the results for all datasets are in Table~\ref{tab:results}. \par

\begin{table}
	\centering
	\caption{Fundamental parameters of HD 141943.}
	\label{tab:HD141943}
    \begin{tabular}{lc}
    \hline
    Parameter & HD 141943  \\
    \hline
    Spectral type & G2V  \\
    Distance (pc) &  60.028 $\pm$ 0.083 \textsuperscript{d}\\
    Age (Myr) & 17-32\textsuperscript{b}  \\
    $M_{\mathrm{\star}}$ (M$_{\odot}$)& 1.3\textsuperscript{a}   \\
    Photospheric temperature $T_{\mathrm{eff}}$ ($K$) & 5850 $\pm$ 100\textsuperscript{a}  \\
    Spot temperature ($K$) & $\approx$ 3950 \textsuperscript{a}  \\
    $R_{\mathrm{eq}}$ (R$_{\odot}$) & 1.5$^{+0.06}_{-0.05}$\textsuperscript{c} \\
    $i$ ($^{\circ}$) & 70 $\pm$ 10\textsuperscript{a}  \\
    $v\sin i$ (\kms) & 35.6 $\pm$ 0.7 \textsuperscript{e}   \\
    Equatorial rotation period $P_{\mathrm{eq}}$ (d) & $2.198\pm 0.002$\textsuperscript{e}  \\
    $\mathrm{d}\Omega$ (\radd) & $0.1331^{+0.0095\textsuperscript{e}}_{-0.0094}$ \\
    \hline
    \multicolumn{2}{l}{\textsuperscript{a}\footnotesize{M11A}}\\
    \multicolumn{2}{l}{\textsuperscript{b}\footnotesize{\cite{Hillenbrand2008}}}\\
    \multicolumn{2}{l}{\textsuperscript{c}\footnotesize{Gaia DR2: \cite{Gaia2016,Gaia2018}}}\\
    \multicolumn{2}{l}{\textsuperscript{d}\footnotesize{Gaia EDR3: \cite{Gaia2016,Gaia2020}}}\\
    \multicolumn{2}{l}{\textsuperscript{e}\footnotesize{This study}}
    \end{tabular}
\end{table}

\begin{figure*}
    \centering
    \includegraphics[width=\columnwidth]{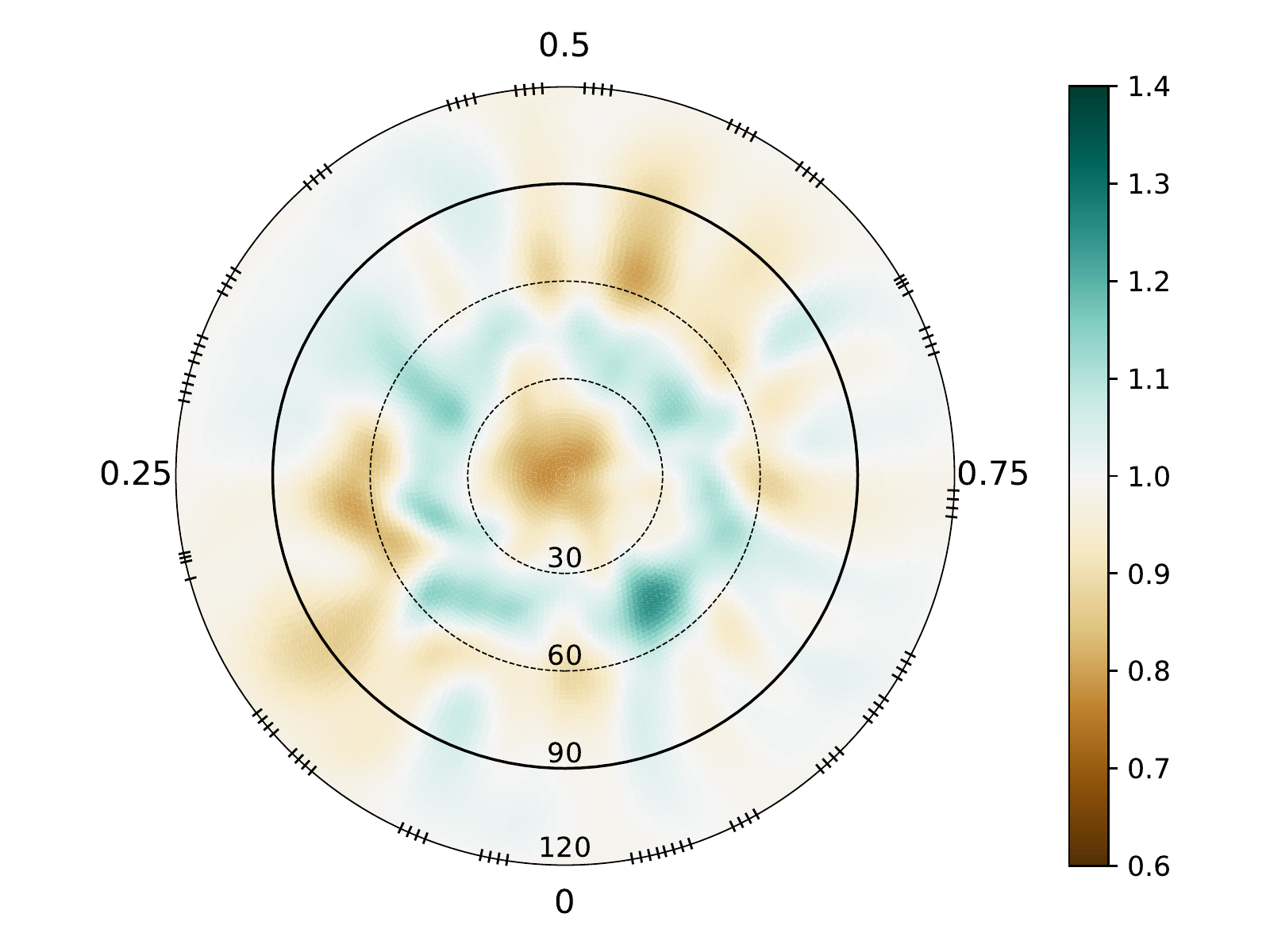}
    \includegraphics[width=\columnwidth]{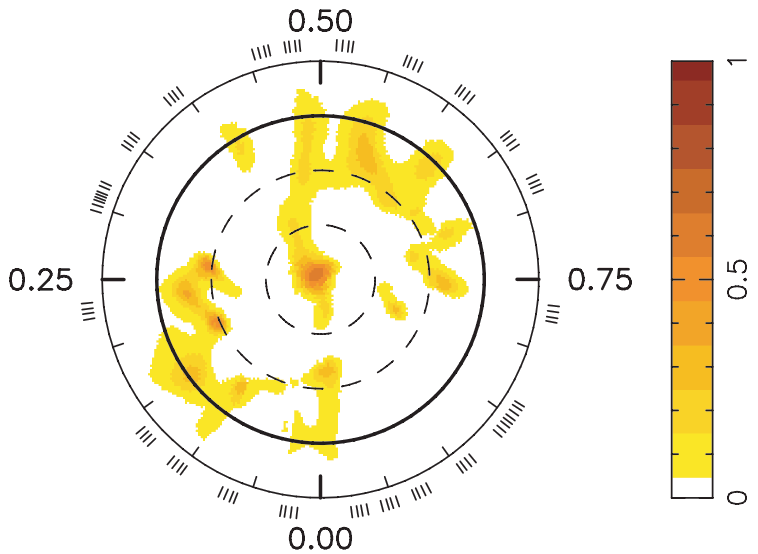}
    \includegraphics[width=\columnwidth]{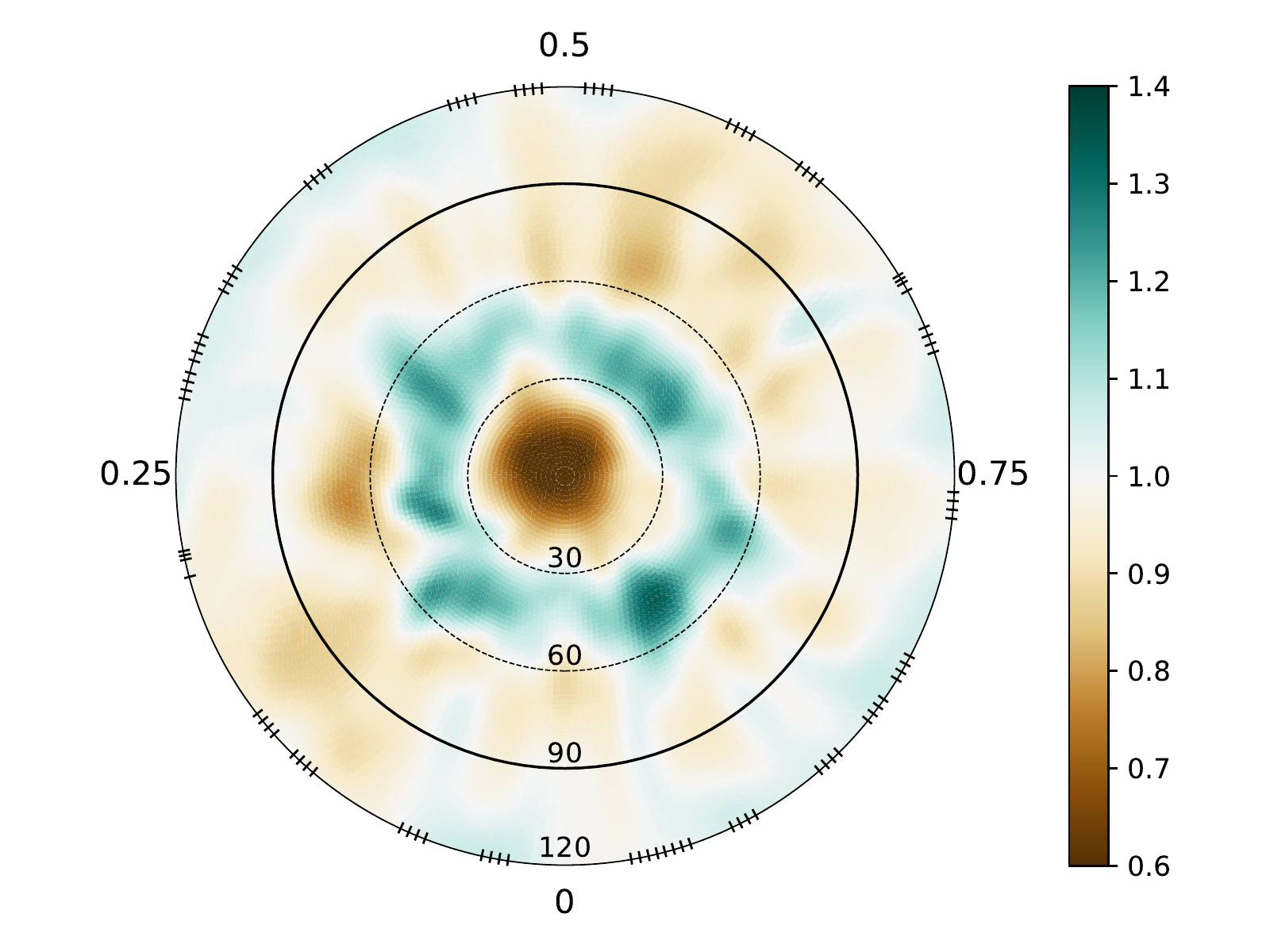}
    \includegraphics[width=\columnwidth]{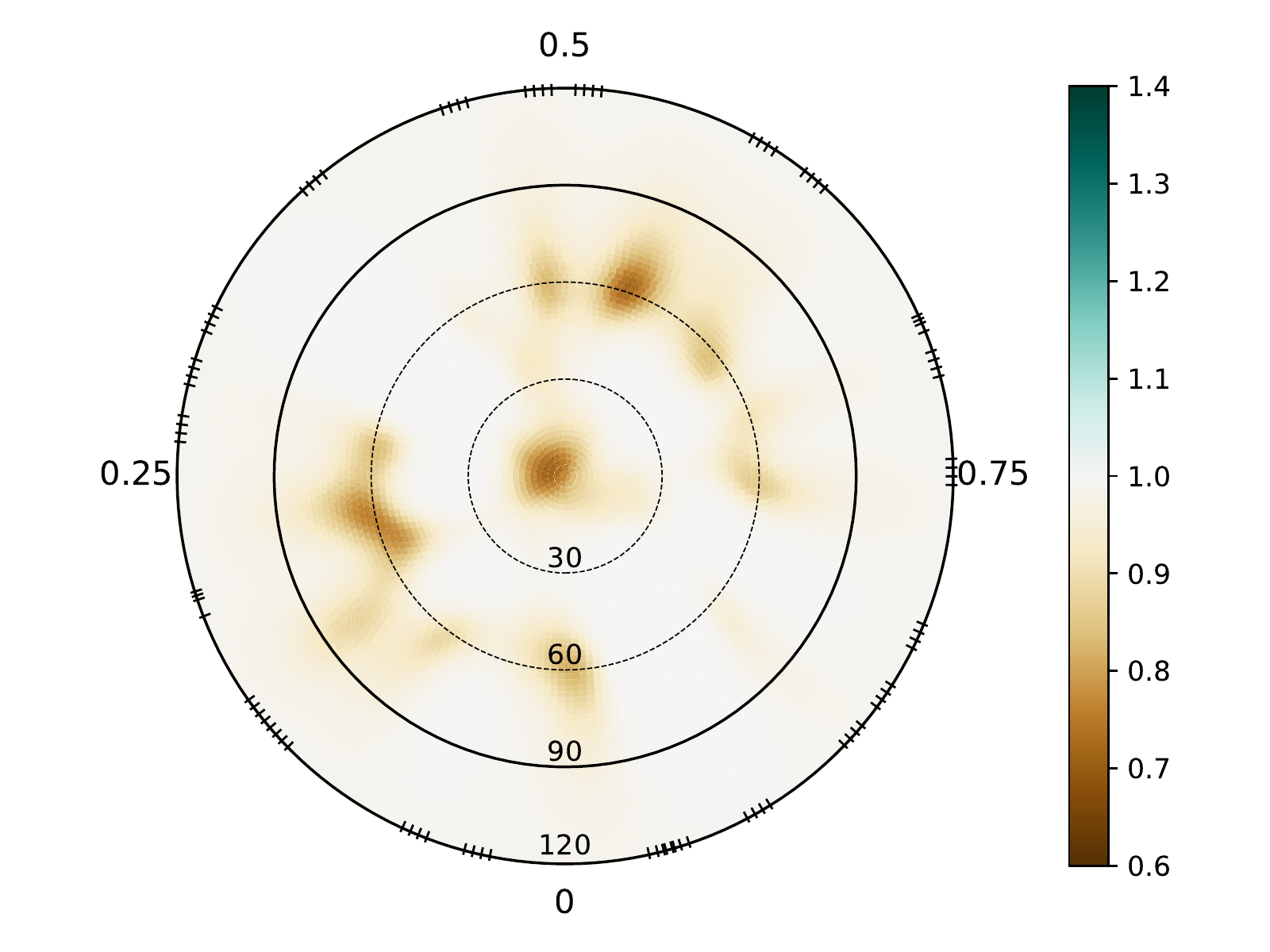}
    \caption{Comparison of the surface brightness maps for HD 141943's original dataset (\#5). Each map is a maximum entropy reconstructed image of the brightness features at the surface of the star. The blue and brown patches indicate regions that are warmer or colder than the photosphere, respectively. The maps are polar projected, with the centre being the visible rotation pole and the full ring labelled 90 the equator. Values on the outermost ring give the rotational phase and the ticks indicate the phase of each observations. Top left:  Map obtained using the parameters from the first line in Table~\ref{tab:bestparam} and mapping for bright and dark features. Top right: Map extracted from M11A (second line of Table~\ref{tab:bestparam}) and mapping only dark features. In their approach, the colour scale expresses the spot filling factor. Bottom left: Map obtained using the parameters from the third line of Table~\ref{tab:bestparam}, with a forced inclination parameter of 70$^{\circ}$ and mapping for bright and dark features. Bottom right: Map obtained using the parameters from the fourth line of Table~\ref{tab:bestparam} and mapping only dark features.}
    \label{fig:maps_comparison}
\end{figure*}

\section{Analysis of HD 141943}
\label{sec:maindataset}

Before attempting to recover injected planets behind HD141943's activity, we analysed the raw observations (dataset \#5 containing no planet) to recover stellar parameters and ensure the star does not host any planet \emph{that we can detect}.

\subsection{Spectropolarimetric dataset}
\label{sec:spectrodataset}
Spectroscopic Stokes \emph{I} (Intensity) and \emph{V} (polarised) observations of HD 141943 used in this study were aquired using the SEMPOL  instrument, visitor polarimeter operating together with the University College London Echelle Spectrograph \citep{Donati2003} and mounted on the 3.9m Anglo-Australian Telescope (AAT) in Siding Spring, Australia. Available data comprises 92 spectra spread over 11 days between March 30 and April 09, 2007, covering 4.68 stellar revolutions, offering a well-sampled rotational phase coverage as required for DI and ZDI (further details on the data can be found in \citealt{Marsden2011a}) and a suitable timescale to search for hot Jupiters. The 92 spectra were taken in chunks of four 30 minute consecutive exposures, each in different polarisation states to perform ZDI. As each 2 hour observing run represents a very short time frame compared to $P_{\mathrm{eq}}$ (stellar equatorial rotation period) and any simulated hot Jupiters' orbital period, this dataset can be treated as containing 23 epochs rather than 92. A previous DI and ZDI analysis of this dataset is in available in \citealt{Marsden2011a,Marsden2011b} (M11A/B hereafter).
Reduction of raw spectra  was done using the \textsc{esprit} pipeline \citep[Echelle SPectra Reduction: an Interactive Tool,][]{Donati1997}.
% The procedure done to translate reduced spectra to radial velocity values was described in section ~\ref{sec:spectratoline} and ~\ref{subsec:RV}.
%%Put this in??
% {\color{red}For all plots in this paper, time is defined each stars, we define the rotational phase as:
% \begin{equation}
%     \phi_{\star} =\frac{\mathrm{BJD} - \mathrm{BJD_{{\star}_{mid}}}}{P_{eq}}	\label{eq:planet}
% \end{equation}
% Where BJD, $\mathrm{BJD_{{\star}_{mid}}}$ and $P_{\star}$ are, respectively, the  Barycentric Julian Date in Barycentric Dynamical Time \citep{Eastman2010}, the BJD of the mid-point of the observations in the data set and the rotation period (see Table {~\ref{tab:meanhJD}}).}

\subsection{Stellar parameters and surface mapping}
\label{subsec:refinedstellarparam}

\begin{table}
	\centering
	\caption{Set of parameters resulting from 4 different analysis: (i) Bright and dark features mapping, (ii) from M11A, (iii) bright and dark features mapping with an inclination angle constrained to 70$^{\circ}$ matching M11A's value and (iv) dark features only.}
	\label{tab:bestparam}
    \begin{tabular}{lllll}
    \hline
    Best value & $v$ $\mathrm{sin}$ $i$ (\kms) & $P_{\mathrm{eq}}$ (d) & $\mathrm{d\Omega}$ (\radd) & $i$ ($^{\circ}$) \\
    \hline
    This work & 35.6 & 2.198 & 0.13 & 43 \\
    M11A/B & 35 & 2.182 & 0.36 & 70 \\
    Fixed $i$ & 35.6 & 2.197 & 0.12 & 70 (fixed) \\
    Dark only & 35.4 & 2.214 & 0.02 & 42 \\
    \hline
    \end{tabular}
\end{table}
HD 141943 is a young ($\approx$ 17-32 Myr, M11A and \citealt{Hillenbrand2008}), nearby (60 $\pm$ 0.08 pc, estimated using \citealt{vo:eDR3_lite_dist} with Gaia EDR3 data \citealt{Gaia2016,Gaia2020}) active G pre-main sequence star. This Sun like star has a mass of 1.3 M$_{\odot}$ and a radius of 1.5 $^{+0.06}_{-0.05} \ \mathrm{R_{\odot}}$ (Gaia DR2, \citealt{Gaia2018}). \cite{Soummer2014} also identified a surrounding near edge-on debris disk, consistent with a planetesimal belt populated by two dust components at respective grain temperatures of 60 $K$ and 202 $K$. The extended list of stellar parameters can be found in Table~\ref{tab:HD141943}. \par 
We inferred stellar parameters by analysing the raw HD 141943 dataset, containing no injected planet. These are marked with the superscript \textsuperscript{d} in Table~\ref{tab:HD141943}: $v$ $\mathrm{sin}$ $i$ = 35.6 $\pm$ 0.7\,\kms, $i$ = 43 $\pm$ 10$^{\circ}$, $P_{\mathrm{eq}}$ = 2.198 $\pm$ 0.002\,d and $\mathrm{d}\Omega$ = $0.1331^{+0.0095}_{-0.0094}$\,\radd. \par
These parameters are close although not exactly matching the previous analysis from M11A/B (see the first 2 lines from Table~\ref{tab:bestparam}). This discrepancy could be explained by the fact that the DI/ZDI code used between M11A/B is slightly different than ours. Mainly, \textsc{zdipy} let us map bright and dark surface features (spots) in contrast with only dark spots in M11A/B. The inclination is the parameter with the largest difference (43$^{\circ}$ vs 70$^{\circ}$), and also the hardest to constrain. To further investigate, we derived the best solution when fitting both (i) only for dark spots and (ii) using both dark and bright spots but forcing $i$ to match M11A/B's value (i.e. 70$^{\circ}$). Obtained Doppler maps and best parameters for the three cases (dark + bright, only dark and dark + bright with imposed 70$^{\circ}$ inclination) are given in Figure~\ref{fig:maps_comparison} and Table~\ref{tab:bestparam}, respectively.\par
These three cases yielded similar results, however noting the negligible differential rotation when fitting only the dark features. Forcing the $i$ to 70$^{\circ}$ did not change the overall solution, and we found good agreement between the dark + bright non forced and forced analysis. The contrast difference on the Doppler maps as seen on the bottom-left map of Figure~\ref{fig:maps_comparison} is due to the effect of projection imposed by $i$. Spot locations are consistent across all maps and with M11A. The difficulty to constrain the inclination angle prevents a reliable deduction of the stellar radius $\mathrm{R_{eq}}$ and we, therefore, used the Gaia DR2's value given in Table~\ref{tab:HD141943}.
Our main objective for this paper was to filter out as efficiently as possible any rotationally-modulated signal in RVs. Since setting $i$ to 43$^{\circ}$ optimises this task, we adopted this value for the inclination in the rest of this study.

\par
\begin{figure}
    \centering
	\includegraphics[width=\columnwidth]{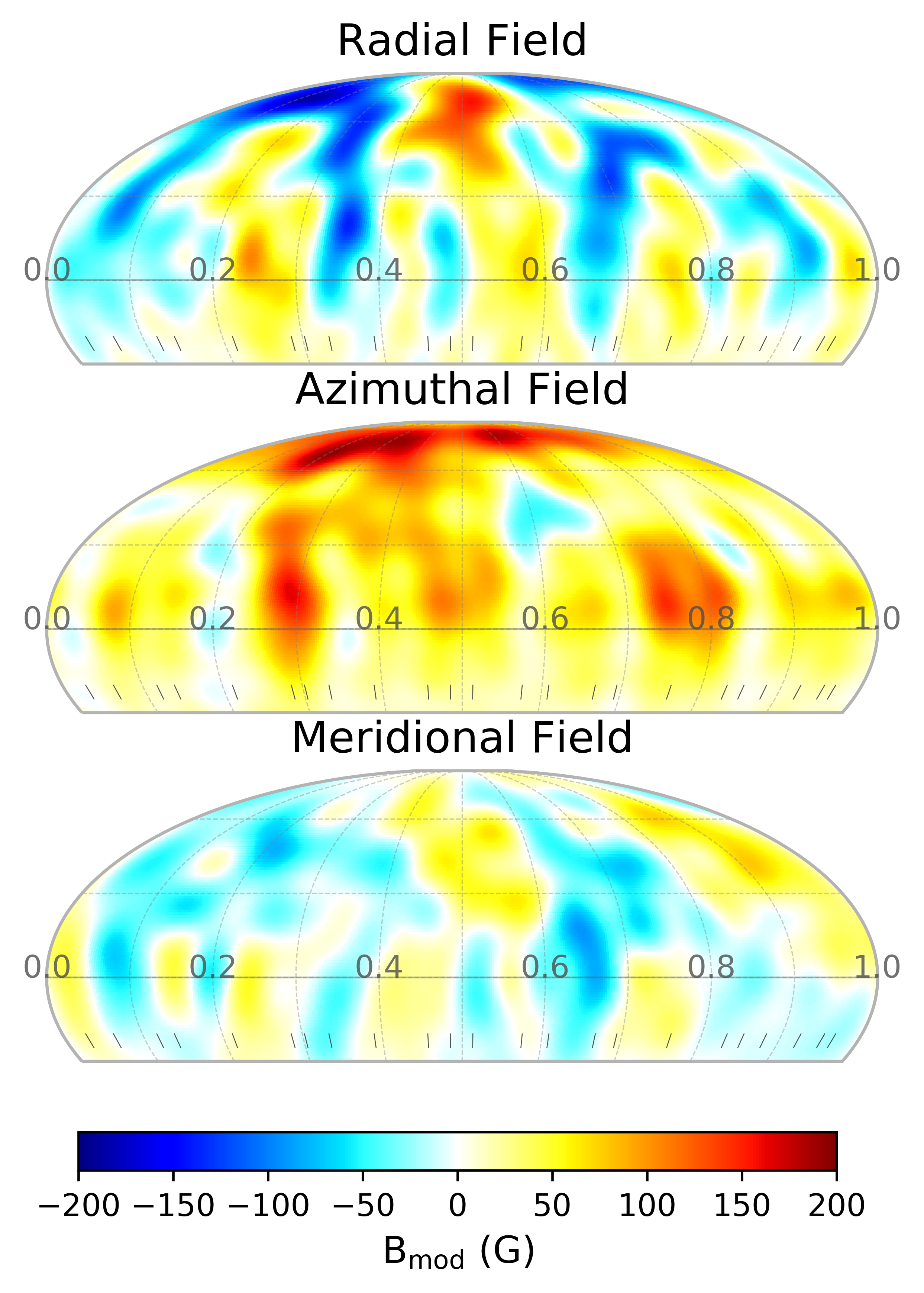}
    \caption{Maximum entropy image reconstruction of the radial (top), azimuthal (middle), and meridional (bottom) components of the magnetic field for HD 141943. Positive/negative field modulus values (in Gauss) are displayed in yellow/red and blue respectively. The horizontal line shows the equator with the number describing the phase. Ticks translate each measurement's epoch.}
    \label{fig:mag_map}
\end{figure}

Figure~\ref{fig:mag_map} shows the radial (top), azimuthal (middle), and meridional (bottom) magnetic field distribution, derived with ZDI using Stokes \emph{V} LSD profiles. We find a magnetic field with 52 and 48 per cent distribution for the poloidal and toroidal components respectively, well agreeing with the 47 and 52 per cent from M11A. The mean strength is $\approx$ 52 Gauss, much lower than M11A's value of 91 G. This can be explained again by the difference in inclination angle. Indeed, re-applying ZDI with a forced $i$ = 70$^{\circ}$ yields a field strength of 85 G, better agreeing with M11A.

\subsection{Planet search}
\label{subsec:Planet search}
Before injecting planets in the HD 141943 dataset, we ensured 
it did not exhibit any sign of hosting a planet. \par
\begin{figure*}
    \centering
    \includegraphics[width=0.8\textwidth]{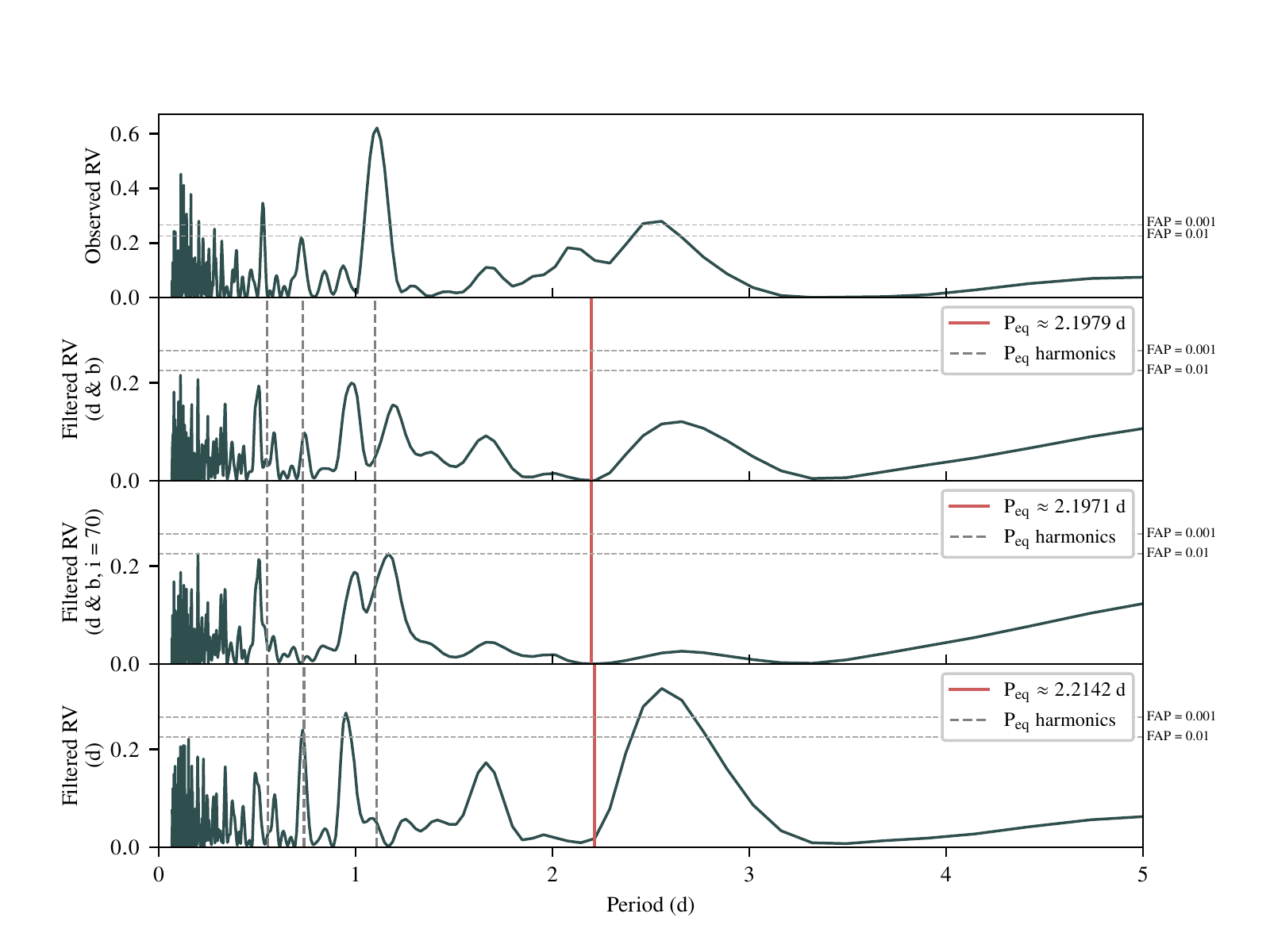}
    \caption{Lomb-scargle periodograms for the original dataset (\#5, containing no planet). 1st panel: Observed (raw) RVs. All other panels are for filtered RVs (i.e. raw RVs - synthetic RVs, where synthetic RVs are derived from the Doppler Imaging fitting). 2nd panel: Filtered RVs using the dark and bright (d \& b) features for DI. 3rd panel: Filtered RVs using the dark and bright features for DI and with the 70$^{\circ}$ constrain on $i$  (d \& b, $i$ = 70). 4th panel: Filtered RVs using only dark features (d) for DI. We note that the peak around 2.6 days is likely caused by the rotation period at intermediate latitudes (offering a maximal visibility given the inclination angle of the star). This value is larger than the equatorial period depicted by the red, vertical lines, as expected in the case of a differentially rotating surface (see Section~\ref{subsec:diffrot}).}
    \label{fig:per_5}
\end{figure*}

 
The top panel of Figure~\ref{fig:per_5} shows the periodogram of the raw RVs, where we identified $P_{\mathrm{eq}}$ and its harmonics, the strongest signature being present at $P_{\mathrm{eq}}/2$. Second, third and fourth panels are periodograms of the filtered RVs, respectively from dark and bright, dark and bright with imposed $i = 70^{\circ}$ features and only dark analysis. All show similar features but one peak (around 2.7 days) did show different height across analysis, and was above the 0.001 FAP threshold for the dark spot only analysis. However, it did not reach overwhelming significance. This dataset did not allow us to assess the impact of the varying DI solutions  (dark, dark + bright and dark + bright with forced inclination) on the planet retrieval as it has no injected planet. To test that, we performed a second analysis using these three configurations for dataset 22 (see section \ref{sec:simulations} for details on simulated datasets), containing a simulated planet in the `uncertain' range of detection. We found that the different DI solutions did not change our conclusions regarding the planet search (detailed analysis available in Appendix~\ref{sec:appendixC}).

The GP confirmed that we were not able to detect any significant planet in the raw dataset. We found a Bayes Factor in favour of the single planet model over the activity only model of only 0.3 (p $\approx$ 0.23).

\begin{figure*}
	\includegraphics[width=\textwidth]{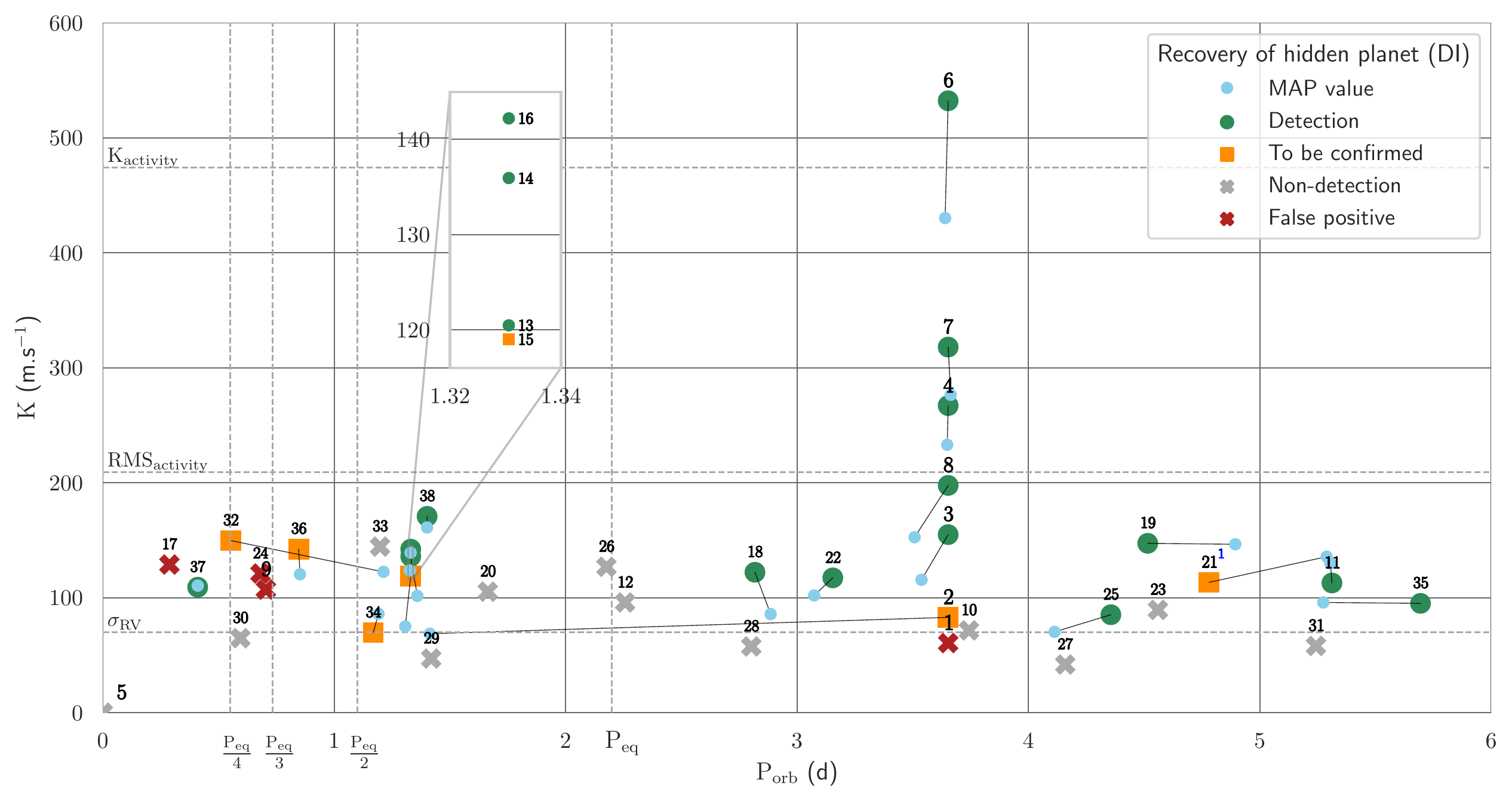}
    \caption{Results using the Doppler Imaging method. Each marker on the plot is a dataset containing a simulated injected planet. Orbital period (d) is on the x-axis and semi-amplitude (\mos) is on the y-axis. Green circles: identification of the correct planet with a periodogram peak above a FAP of 0.001. Orange square: two peaks of similar height (see Section~\ref{fig:result_DI}) were found preventing a safe conclusion or when the width of the correctly identified peak yielded a deviation of more than 10 per cent between the retrieved and injected orbital period. Grey crosses: no signature with FAP > 0.001 could be identified. Red crosses: the most significant peak was not matching the injected period peak. Horizontal dashed lines show the stellar activity semi-amplitude ($K_{\mathrm{activity}}$) and the error bars on the retrieved RVs ($\mathrm{\sigma_{RV}}$). Vertical dashed lines show the rotation period of the star and its harmonics.Blue points show, for the `Detection' and `To be confirmed' datasets, the $P_{\mathrm{eq}}$ and $K$ values corresponding to the highest peak in the periodogram.\textsuperscript{1}: For dataset \#21, the correct peak was found, at a FAP of $1.1\times10^{-13}$. However, that peak  being very broad, it yielded a 10.6 per cent deviation between the retrieved and the injected period, slightly outside our 10 per cent limit.}
    \label{fig:result_DI}
\end{figure*}

\section{simulated exoplanet datasets}
\label{sec:simulations}

We created 37 datasets, each containing a single planet on a circular orbit around HD 141943, following a procedure described in~\ref{subsec:Planets} and \ref{subsec:completedatasets}. Each planet was incorporated into the raw spectra studied in the previous section.

\subsection{Injected Planets}
\label{subsec:Planets}
Injected planets were chosen to be massive short-period exoplanets, with masses ranging from 0.38 up to 5.9\,$\mathrm{M_{J}}$ and periods shorter than 6 days. We set the orbits to be circular, as it is believed to be the case for most HJs, especially for orbits shorter than 3 days \citep{Dawson2018}. We should nonetheless bear in mind that eccentricity can be a crucial aspect in favour of high-eccentricity migration and should not be overlooked especially when attempting to detect the slightly cooler warm Jupiters ($P_{\mathrm{orb}}$ > 10 days). The RV shift induced by each planet was defined as:
\begin{equation}
    RV_{\mathrm{pla}}(t) = K\sin \left[2\pi\left(\frac{t}{P_{\mathrm{orb}}} - \Phi + 0.5 \right)\right]
	\label{eq:planet}
\end{equation}
with $K$ the semi-amplitude of the signal, $P_{\mathrm{orb}}$ the planet's orbital period and $\Phi$ the phase. $\mathrm{\Phi} \in \left[ 0:1\right]$ and was defined such that when $\mathrm{\Phi} = 0$,  the planet crosses the plane containing the line of sight. We set $\mathrm{\Phi} = 0$ to match the mid-point of the observations ($BJD_{{\mathrm{\star}_{mid}}}$ = 2454195.153776). 
For the rest of this paper, we will refer to semi-amplitude values ($K$) for the planets rather than mass. The equivalence between $K$ and mass is described in Section~\ref{subsec:recoveredexoplanets}.

\subsection{Complete datasets}
\label{subsec:completedatasets}
To build our datasets, we generated an RV time series using Equation~\ref{eq:planet} at times matching our observing epochs and then shifted each spectrum accordingly in wavelength space. In order to explore our planetary parameter space (made of $K$, $P_{\mathrm{orb}}$ and $\Phi$) without being overwhelmed with the number of datasets to analyse ($\mathrm{n_{K}} \times \mathrm{n_{P_{orb}}} \times \mathrm{n_{\Phi}}$), we used the following strategy: \par
First, we created 7 datasets (\#1 to \#8, excluding \#5, the original one) at a fixed period (3.653 days) with $K$ ranging from 50 to 500 $\mathrm{m.s}^{-1}$ and random $\Phi$. This initial analysis provided an estimate of the limiting semi-amplitude range for detectability. \par
Then we generated additional datasets 4-5 at a time, filling areas in the parameter space that seemed relevant, i.e. around the noise limit, around $P_{\mathrm{eq}}$ and harmonics and to cover empty areas of the parameter space. After generating each batch of datasets, these were randomly assigned a mock value before analysis to avoid biases. \par
We ended up with 37 datasets (38 when including the original dataset) spanning the following ranges: 42 to 532\,\mos in $K$, 0.288 to 5.69 days for $P_{\mathrm{orb}}$ and 0.02 to 0.99 in $\Phi$. See Table~\ref{tab:results} for specific details.

\subsection{Radial velocity extraction}
\label{subsec:RV_extractin}
We tested the RV extraction from the line profiles using both a first order moment approach and a Generalised Normal distribution fit. For the original dataset (\#5 with no injected planet) and given our limits on the integration for the FOM approach the average difference between FOM and GND extracted RVs (from the observed profiles) is 8\,\mos with a maximum difference of 144\,\mos for the most extreme point. The uncertainty on the RVs also differed as GND yielded uncertainties twice as large as those from FOM (148.5$\pm$2.0\,\mos vs 70.4$\pm$1.8\,\mos). \par
As previously mentionned, FOM derived RVs and uncertainties are dependent on the number of points taken into account for its computation (i.e. chosen integration limits for the line profile) and where to cut in the wings of the line profile can be somewhat arbitrary. On both sides of each profiles (and for all of them), we cut at 43\,\kms relative to the line center. We tested different limits and chose the one that gave the least average difference with the GND approach, which is an analytical function and therefore not prone to this effect.\par

\section{Results}
\label{sec:Results}

All datasets are extensively described in Table~\ref{tab:results} and  will be referred to using their number (\#1, \#2, ...,\#38).

\subsection{Stellar parameters}
Stellar parameters inferred from the datasets containing injected planets are consistent with our refined parameters derived from the raw dataset (\#5, see Section~\ref{subsec:refinedstellarparam}). The mean of the retrieved stellar parameters across all datasets, along with the largest deviation from the mean (given by the $\pm$) are: $v\sin i$ = 35.6 $^{+0.27}_{-0.45}$ \kms, $i$ = 43 $^{+5}_{-3}$ $^\circ$, $P_{\mathrm{eq}}$ = 2.1995 $^{+0.007}_{-0.006}$ days and $\mathrm{d\Omega}$ = 0.119 $\pm 0.08$ \radd. In all cases, spot distributions are similar, with slight differences in terms of contrast. This can be explained by the fact that the fit to the line profiles was sometimes performed to a slightly different $\chi^2$ level. Typically, the presence of a planet with a semi-amplitude significantly larger than the activity level (e.g. for dataset \#6) slightly impacts the performance of the DI. However, even such a large planet signature did not hamper the capacity of the DI to identify spot locations and recover the planet. \par

\begin{figure*}
	\includegraphics[width=\textwidth]{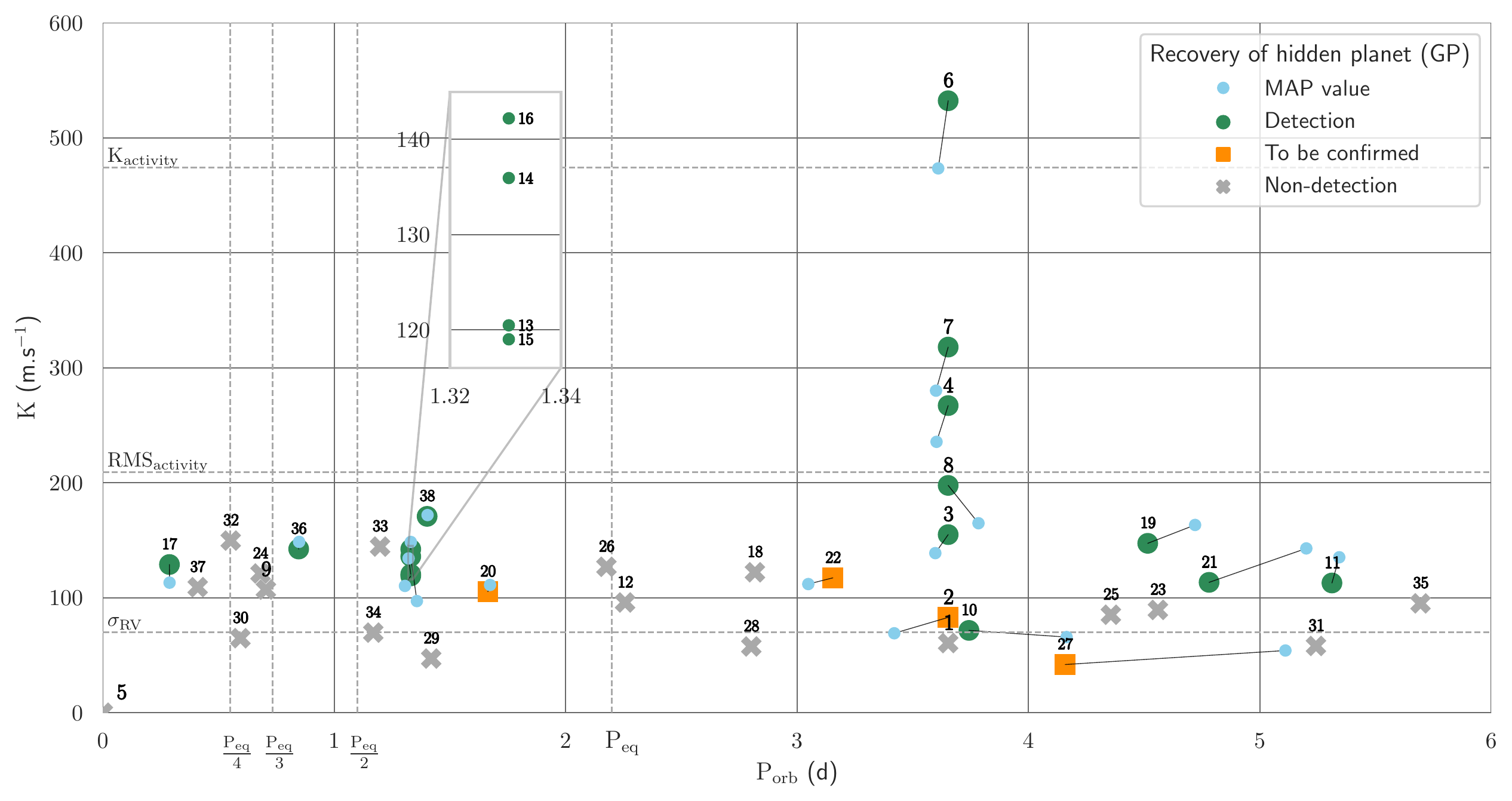}
    \caption{Results using the Gaussian Process method. Each marker on the plot is a simulated injected planet. Orbital period (d) is on the x-axis and semi-amplitude (\mos) is on the y-axis. Green circles: the evidence between the single-planet model and with activity only is $\Delta$ ln($\mathcal{Z}$) > 10, associated with at least a strong detection (probability p > 0.909). Orange squares: 10 > $\Delta$ ln($\mathcal{Z}$) > 3 or substantial evidence (0.95 > p > 0.75). Grey crosses: $\Delta$ ln($\mathcal{Z}$) < 3 (p < 0.75). Horizontal dashed lines show the stellar activity semi-amplitude ($K_{\mathrm{activity}}$) and the error bars on the retrieved RVs ($\sigma_{{RV}}$). Vertical dashed lines show the rotation period of the star and its harmonics. Blue points show the Maximum A Posteriori (MAP) values for $P_{\mathrm{eq}}$ and $K$, linked by a line to the corresponding dataset.} 
    \label{fig:result_GP}
\end{figure*}

% \begin{figure*}
% 	\includegraphics[width=\textwidth]{plots/result_method_comparison.pdf}
%     \caption{Results combining both techniques (DI + GP). Each marker represents a simulated injected} planet. Orbital period (d) is on the x-axis and semi-amplitude (\mos) is on the y-axis. Dark green circles: positive detections. Orange squares: detected signature requiring further observation to be confirmed. Grey crosses: non-detections. Light green circles and gold squares indicate dataset for which the combined use of both techniques acted to conclude differently than if only on method would have been used. Light green circles: detection confirmed/dictated by the GP. Gold squares: DI acted to increase the dataset status to `requiring further confirmation' when it was a non-detection for the GP. Horizontal dashed lines show the stellar activity semi-amplitude ($K_{\mathrm{activity}}$) and the error bars on the retrieved RVs  ($\mathrm{\sigma_{RV}}$). Vertical dashed lines show the rotation period of the star and its harmonics.}
%     \label{fig:result_method_comparison}
% \end{figure*}

\subsection{Planet detection: Methods performance}

\subsubsection{Method 1: Doppler Imaging activity filtering}
\label{subsec:resultDI}

Results are shown in Figure~\ref{fig:result_DI}. Each marker represents a dataset with its corresponding number in order to easily refer to Table~\ref{tab:results} containing more details. Marker positions indicate the injected planets' $K$ and $P_{\mathrm{orb}}$ (as we did not identify a systematic impact of $\Phi$ on retrievals, it was omitted for clarity). Each dataset is identified by a specific marker/colour combination: A green circle when a periodogram peak was identified above a FAP threshold of 0.001 (0.1 per cent) and with a deviation from the true period of $< 10$ per cent, an orange square when two peaks above a FAP of 0.001 and of similar height were found or when the right peak was identified but with a deviation from the true period of $> 10$ per cent, preventing a safe conclusion, a grey cross when no signature above FAP = 0.001 could be identified and finally a red cross when a peak was present above the FAP threshold but was not matching the injected planetary period, i.e. a false positive. \par 
This approach yielded 16 positive detections, 6 inconclusive findings, 11 non-detections and 4 false positives. These 4 datasets confirm that using FAP as a measure of significance is not the safest approach, as discussed in Section~\ref{subsec:DI filtering}. Rigorous estimation of the significance was performed with the GP analysis. \par
All 6 simulated planets with semi-amplitudes larger than 150\,\mos\, were well retrieved and half (8/18) between 100\,\mos\, and 150\,\mos. This fraction increased to 60 per cent (8/13) when removing all datasets close to $P_{\mathrm{eq}}$ and its harmonics. Only 1/13 planets below 100\,\mos\, were found (noting that \#25 was a very weak detection with FAP $= 3.5\times10^{-4}$). \par
We note that all analysis that identified the right peak but with deviation of more than 5 per cent (up to 10.6 per cent for \#21) from the true period (\#19, \#21, \#25 and \#35) had fewer than 2.5 orbital periods within our observation time span. This inaccuracy was due to the width of the peaks in the periodogram arising as the period represented a significant fraction of the time span. It is safest, given our number of samples, to cover at least 2 to 3 orbital periods to achieve sufficient precision on $P_{\mathrm{orb}}$. \par
The 6 `to be confirmed' (orange squares) datasets: \#2, \#15, \#32, \#34 and \#36 all exhibited two competing peaks above the FAP threshold and at a similar height (FAP of $1.8\times10^{-4}$ \& $7.2\times10^{-4}$ for \#2, $1.6\times10^{-4}$ \& $1.4\times10^{-4}$ for \#15, $1.1\times10^{-11}$ \& $1.1\times10^{-11}$ for \#32, $1.25\times10^{-6}$ \& $2.6\times10^{-4}$ for \#34 and $1.15\times10^{-10}$ \& $2.43\times10^{-9}$ for \#36) preventing us from being able to choose the correct period. For \#2, \#15 and \#32, the peaks are just above our detection threshold and it is therefore not surprising to find competing features. For \#32 and \#36 however, the competing peaks were both very significant. We are not sure as to how to interpret these, which vanished as we filter the signature associated with one of the two peaks. As shown in Figure~\ref{fig:periodograms32_36}, these spurious peaks did not seem to correspond to any harmonics of neither the planet nor the star. See periodograms. Although complex interactions between the uneven data sampling and the periodic signatures cannot be ruled out, no significant peak could be identified in the window function (see Figure~\ref{fig:window}). The complete analysis of these 2 datasets can be found in Appendix~\ref{sec:appendixB}. Dataset \#21 also falls in the 'orange square' category with a very wide identified peak, yielding a 10.6 per cent deviation between the retrieved and injected period, slightly over our 10 per cent threshold. \par 
False positives arose as the highest peak was not the simulated one, which would lead to false identification (if relying solely on DI) for \#1 and \#17. Regarding \#9 and \#24, the peaks were barely above the FAP of 0.001 and would not have led to a significant detection. \par
Although we could not identify a systematic impact, phase is expected to play a role in the injected planets retrieval and we can see this occurring in the zoomed box in Figure~\ref{fig:result_DI}. The only noticeable difference between \#13 and \#15 is their phases (respectively $\Phi_{13}$ = 0.4769 and $\Phi_{15}$ = 0.1093) and yet planet \#13 is recovered but not \#15. \par
Studying periodograms for all datasets indicated that planets with periods close to $P_{\mathrm{eq}}$ (\#12 and \#26), $P_{\mathrm{eq}/2}$ (\#33 and \#34), $P_{\mathrm{eq}/3}$ (\#9 and \#24) and  $P_{\mathrm{eq}/4}$ (\#30 and \#32) seem to be affected by the activity filtering. The case of \#32 has been discussed above. This effect is to some extent expected as DI has the capacity to distort the line profiles, interpreting the rotationally modulated distortions as produced by spots on the brightness maps at harmonics of the rotation period and therefore is likely to absorb part of a planetary signature close to one of these periods. \par
For RV searches, the LS periodogram has limitations (choosing a FAP limit, interpreting the significance of a result, limitation to sinusoidal signals, see \citealt{Vanderplas2018}) and we emphasize that dedicated treatment for stellar activity should be performed. We therefore advocate for incorporating a second, complementary method, presented in the following section, allowing both better quantification of the significance of a retrieved signature and a more comprehensive modelling of the activity. \par

%%%%%%%%%%%%%%%%%%%%%%%%%%%%%%%%%%%%%%%%%%%%%%%%%%
% TABLE PART 1  
%%%%%%%%%%%%%%%'%%%%%%%%%%%%%%%%%%%%%%%%%%%%%%%%%%%

\begin{table*}
\scriptsize
\centering
	\caption{Datasets 1 to 6. Each column represents a simulated dataset. The first section (rows 1-4) are the stellar parameters inferred from the Doppler Imaging analysis. The second section (rows 5-7) gives the values of the three parameters used to simulate the injected planet. The third section (rows 8-13) gives the results of the Method \#1 DI filtering. $RV_0$ is an offset, then we have the three recovered planet parameters, followed by the rms and $\chi^2$ of the residuals. The fourth section (rows 14-22) are the result of the GP (Method \#2) for the no planet (activity only) model. For the parameters ($\theta_1$ to $\sigma_s$), the values are given as: mean $\pm$ std (maximum a posteriori). We then have the rms and $\chi^2$ of the residuals and the resulting natural logarithm of the evidence. The last section (rows 23-36) are the result of the GP for the single planet model. Again, the parameters ($\theta_1$ to $\sigma_s$) values are given as: mean $\pm$ std (maximum a posteriori). We then have the rms and $\chi^2$ of the residuals and the resulting natural logarithm of the evidence. Finally, we give the Bayes Factor (BF), defined as the ratio between $\mathcal{Z}$ from the single planet model (row 34) and the no planet model (row 22). The last row is the probability in favour of the single planet model associated with the BF value. Only the first 6 rows (datasets) are shown here, the full version is available as online material.}
	\label{tab:results}
    \begin{tabular}{lccccccc}
    \hline
    Dataset & \#1 & \#2 & \#3 & \#4 & \#5 & \#6 & ...\\
    \hline
      \multicolumn{7}{l}{Doppler Imaging inferred stellar parameters}\\
      $i$($^{\circ}$) & 43 & 40.5 & 41.5 & 42 & 42.5 & 45 & ...\\
      $v\sin i$ (\kms) & 35.47 & 35.53 & 35.75 & 35.789 & 35.643 & 35.862 & ...\\
      $P_{\mathrm{eq}}$   (d) & 2.20615 & 2.20345 & 2.1988 & 2.19408 & 2.19788 & 2.20179 & ...\\
      $\mathrm{d}\Omega$   (\radd) & 0.0769 & 0.09796 & 0.13333 & 0.14694 & 0.13333 & 0.12308 & ...\\
     \hline
      \multicolumn{7}{l}{Injected planet parameters}\\
      \\  
    $K$ (\mos) & 60.6 & 82.9 & 154.9 & 267.1 & - & 532.2 & ...\\
    $P_{\mathrm{orb}}$ (d) & 3.6531 & 3.6531 & 3.6531 & 3.6531 & - & 3.6531 & ...\\
    $\Phi$ [0:1] & 0.271 & 0.303 & 0.445 & 0.311 & - & 0.432 & ...\\
    \hline
    \multicolumn{7}{l}{Method \#1 (DI) mean $\pm$ std} \\
    \\
    $RV_\mathrm{0}$ (\mos) & 13.1$\pm$9 & 1.7$\pm$9 & 0.5$\pm$9 & 12.2$\pm$10 & - & -60.0$\pm$10 & ...\\
    $K$ (\mos) & 85.5$\pm$12 & 68.7$\pm$11 & 115.6$\pm$12 & 233.0$\pm$13 & - & 430.2$\pm$14 & ...\\
    $P_{\mathrm{orb}}$ (d) & 2.549$\pm$0.054 & 1.413$\pm$0.022 & 3.538$\pm$0.067 & 3.649$\pm$0.041 & - & 3.640$\pm$0.022 & ...\\
    $\Phi$ [0:1] & 0.612$\pm$0.026 & 0.209$\pm$0.031 & 0.435$\pm$0.018 & 0.301$\pm$0.009 & - & 0.427$\pm$0.006 & ...\\
    rms (\mos) & 81.9 & 79.6 & 79.7 & 84.4 & - & 89.9 & ...\\
    $\chi^2$ & 1.34 & 1.28 & 1.28 & 1.43 & - & 1.63 & ...\\
    \hline
      \multicolumn{7}{l}{Method \#2 (GP) / no planet model / mean $\pm$ std (MAP)} \\
      \\
    $\theta_1$ (\mos) & 314.2$\pm$113.4(217.4) & 276.8$\pm$89.9(208.2) & 265.4$\pm$68.0(217.1) & 320.5$\pm$74.1(277.5) & 357.6$\pm$109.1(296.6) & 462.9$\pm$98.0(394.1) & ...\\
    $\theta_2$ (\mos) & 10.2$\pm$6.3(6.9) & 6.9$\pm$3.8(6.0) & 3.0$\pm$1.4(4.2) & 1.7$\pm$0.5(1.7) & 19.2$\pm$9.3(15.7) & 1.4$\pm$0.3(1.1) & ...\\
    $\theta_3$ (\mos) & 2.190$\pm$0.036(2.148) & 2.190$\pm$0.041(2.141) & 2.205$\pm$0.045(2.164) & 2.206$\pm$0.049(2.255) & 2.215$\pm$0.020(2.216) & 2.207$\pm$0.049(2.268) & ...\\
    $\theta_4$ (\mos) & 0.527$\pm$0.148(0.399) & 0.468$\pm$0.136(0.347) & 0.415$\pm$0.137(0.234) & 0.587$\pm$0.146(0.511) & 0.628$\pm$0.083(0.597) & 0.723$\pm$0.151(0.605) & ...\\
    $RV_\mathrm{0}$ & 12.5$\pm$156.8(-20.5) & 19.7$\pm$127.6(16.6) & 35.6$\pm$97.7(52.1) & 15.8$\pm$113.3(-0.7) & 25.3$\pm$194.0(50.9) & 17.4$\pm$160.3(49.6) & ...\\
    $\sigma_{s}$ & 12.3$\pm$9.5(0.4) & 11.7$\pm$9.1(2.8) & 12.0$\pm$9.2(9.8) & 12.0$\pm$9.2(3.0) & 11.5$\pm$8.8(0.6) & 12.3$\pm$9.5(0.7) & ...\\
    rms (\mos) & 55.9 & 54.9 & 54.8 & 55.8 & 59.0 & 56.6 & ...\\
    $\chi^2$ & 0.62 & 0.59 & 0.59 & 0.61 & 0.69 & 0.63 & ...\\
    $\mathrm{ln{\mathcal{Z}}}$ & -550.99 & -552.15 & -560.46 & -563.85 & -545.03 & -571.93 & ...\\
    \hline
      \multicolumn{7}{l}{{Method \#2 (GP) / single planet model / mean $\pm$ std (MAP)}} \\
      \\
    $\theta_1$ (\mos) & 340.1$\pm$117.7(271.1) & 360.8$\pm$120.8(257.2) & 400.9$\pm$151.9(301.2) & 401.0$\pm$138.3(347.7) & 357.3$\pm$107.9(292.2) & 473.5$\pm$199.5(338.4) & ...\\
    $\theta_2$ (\mos) & 14.0$\pm$7.8(13.3) & 17.2$\pm$10.0(17.1) & 20.9$\pm$13.2(15.3) & 21.7$\pm$12.3(19.0) & 19.9$\pm$9.7(15.9) & 27.7$\pm$16.7(22.3) & ...\\
    $\theta_3$ (\mos) & 2.204$\pm$0.030(2.217) & 2.213$\pm$0.026(2.233) & 2.208$\pm$0.024(2.216) & 2.215$\pm$0.019(2.220) & 2.216$\pm$0.020(2.228) & 2.206$\pm$0.020(2.213) & ...\\
    $\theta_4$ (\mos) & 0.595$\pm$0.151(0.602) & 0.642$\pm$0.154(0.601) & 0.641$\pm$0.163(0.570) & 0.677$\pm$0.143(0.645) & 0.631$\pm$0.081(0.592) & 0.705$\pm$0.156(0.680) & ...\\
    $K$ (\mos) & 43.1$\pm$23.6(59.8) & 59.2$\pm$19.7(69.1) & 130.8$\pm$19.1(138.8) & 238.7$\pm$14.2(235.6) & 29.4$\pm$32.1(39.7) & 474.7$\pm$14.5(473.4) & ...\\
    Period (d) & 3.008$\pm$2.386(3.331) & 3.223$\pm$1.341(3.420) & 3.571$\pm$0.438(3.598) & 3.580$\pm$0.102(3.603) & 2.481$\pm$2.528(0.172) & 3.616$\pm$0.035(3.610) & ...\\
    Phase [0:1] & 0.345$\pm$0.210(0.259) & 0.305$\pm$0.115(0.286) & 0.444$\pm$0.036(0.448) & 0.306$\pm$0.013(0.309) & 0.571$\pm$0.293(0.821) & 0.432$\pm$0.005(0.430) & ...\\
    $RV_\mathrm{0}$ & 38.6$\pm$166.5(-1.6) & 52.6$\pm$178.1(119.5) & 27.8$\pm$212.0(81.8) & 69.8$\pm$211.2(1.2) & 28.4$\pm$190.2(166.2) & 39.2$\pm$292.2(121.2) & ...\\
    $\sigma_{s}$ & 11.0$\pm$8.3(6.8) & 11.1$\pm$8.4(3.9) & 12.4$\pm$9.4(3.2) & 11.8$\pm$9.1(5.1) & 10.7$\pm$8.2(6.3) & 12.7$\pm$9.6(2.8) & ...\\
    rms (\mos) & 57.3 & 62.1 & 61.3 & 57.9 & 60.4 & 64.4 & ...\\
    $\chi^2$ & 0.65 & 0.77 & 0.75 & 0.66 & 0.72 & 0.82 & ...\\
    $\mathrm{ln{\mathcal{Z}}}$ & -551.06 & -550.91 & -556.77 & -556.41 & -546.25 & -560.60 & ...\\
    Bayes Factor & 0.9 & 3.5 & 40.0 & 1702.8 & 0.3 & 83283.0 & ...\\
    p($\mathrm{\mathcal{M}_1}$) & 0.48 & 0.78 & 0.98 & 1.00 & 0.23 & 1.00 & ...\\
    \hline
    \end{tabular}
\end{table*}

%BA: left off here

\subsubsection{Method 2: Gaussian Process regression activity modelling}
\label{subsec:GP}
Again, results are detailed in Table~\ref{tab:results} and summarised in Figure~\ref{fig:result_GP}. We defined successful retrievals (green circles) where the GP strongly favoured the single planet model over the activity only model with a probability p > 0.909 (computed from the marginal likelihood / Bayes factor, see Appendix~\ref{sec:appendixA} and Table~\ref{tab:Bayes factor} for further details). We then have substantial evidence (orange squares) of a planet (i.e. 0.75 < p < 0.909)  and non-detections (grey crosses, i.e. p < 0.75). We note that most of the injected planets (28/37) were correctly identified by the GP, although not always significant enough to lead to a detection claim. \par 
The GP yielded 16 positive detections, 4 `to be confirmed' findings (i.e. requiring further observations), 17 non-detections and more importantly no false positives. Again here, all 6 simulated planets with semi-amplitudes larger than 150\,\mos were correctly found. This drops to half (9/18) between 100\,\mos\, and 150\,\mos (same ratio as DI although not systematically on the same datasets), and increases to 70 per cent (9/13) when removing all datasets close to $P_{\mathrm{eq}}$ and its harmonics. Finally, for planets below 100\,\mos, only 1 out of 13 was found (along with 2 cases requiring further observations, with \#2 a correctly identified planet and \#27 a missed identification). The GP, compared to DI, is more conservative yet more reliable (i.e. no false positives) due to its accurate measure of the significance for each finding. Figure~\ref{fig:result_GP} shows that, similar to the DI analysis, it is difficult to identify planetary signatures close to $P_{\mathrm{eq}}$ or its harmonics. \par
We finally note that imprecision on the retrieved $
P_{\mathrm{orb}}$ increases for longer periods (see MAP values indicated on Figure~\ref{fig:result_GP}). This is because fewer periods are covered by the dataset as we move to the right of Figure~\ref{fig:result_GP}.

\subsubsection{Consistency between methods}
\label{subsec:result_method_comparison}
Utilising two distinct methods serves as cross-validation when a signature is found. However, the GP is the only Bayesian approach, therefore the only one allowing a rigorous quantification of the evidence favouring of a particular model (i.e. presence of a sinusoidal signature in the data or not). \par
For signatures above 150\,\mos and after removing datasets close to $P_{\mathrm{eq}}$ harmonics, both methods yielded systematic detections except for the ambiguity on \#36 when using DI. For signatures between 100\,\mos\, and 150\,\mos, the GP showed more consistency than the DI which exhibited 3 false positives. Out of the 13 datasets below 100\,\mos\, we ended up with 1 detection for both GP and DI. \par
Even though the Bayesian approach using a GP can (i) better handle correlated noise and (ii) more reliably estimate the significance of a detection, the use of the DI filtering method allows an independent validation.

\subsubsection{Comparison with previous work}
In a study performed in 2014, \citet{Jeffers2014} (J14 hereafter) injected various planets behind simulated activity signatures of two young G and K stars. Varying parameters were the planet semi-amplitude, orbital period, and $v\sin i$ (shown to be correlated with the activity level). Stellar activity was generated based on DI maps and with different configurations (e.g. adding plages associated with spots, adding extra random spots, etc. see J14 for more details). The G dwarf was HD 141943, thus making the comparison particularly relevant. Each simulated dataset was composed of a single planet in a circular orbit, to which modelled stellar activity and instrumental signatures were added. In that study, the search for injected planets was performed \emph{without} a specific treatment for stellar activity and was considered successful for periodogram peaks below FAP = 0.01 (vs 0.001 in our study).\par
With 50 observational epochs and for their less complex simulation of activity (only based on the Doppler Imaging maps), J14 were able to retrieve signatures of semi-amplitude $K=110$\,\mos\, when $v\sin i$ = 20\,\kms\, and $K=525$\,\mos\, when $v\sin i$ = 50\,\kms. Regarding their most complex simulation of activity (Doppler Imaging maps + plages + random spots, see J14 for further details) the minimum attainable planetary signature was $K=525$\,\mos\, when $v\sin i$ = 20\,\kms. In the case of $v\sin i$ = 50\,\kms, 200 observational epochs were required to reach the $K=525$\,\mos\, detection threshold. \par 
We note that the data sampling is different which might slightly hinder the comparison \footnote{J14 has one datum per night for 50, 100 or 200 consecutive nights whereas we have 23 epochs over 10 nights.}. With 23 unevenly spread epochs, the smallest signature we could reliably detect was was $K=100$\,\mos\, (down to $70$\,\mos\, for dataset \#10), emphasising the benefit granted by our activity filtering approach. Although now systematically applied by the community for planetary searches around active stars, this emphasises that an independent treatment of stellar activity combined with robust model selection is crucial to improve detection capabilities. \par
%\\
%In \cite{Klein2020} they tested the # of obs, we can't. Around 20 obs is typical for DI/ZDI campaign. # of obs is critical}

\subsubsection{Recovered exoplanets}
\label{subsec:recoveredexoplanets}
Here we translate our results into planetary mass/orbital periods for the case of HD 141943, given $M_{\mathrm{\star}} = 1.3 M_{\mathrm{\odot}}$ and $i$ = 43$^{\circ}$. We consider that the stellar rotation axis is normal to the planet's orbital plan. In this context, our lower detection threshold of 100\,\mos\, is equivalent to a planet with either:
\begin{itemize}
    \item $M_{\mathrm{pla}}$ = 1 $M_{\mathrm{Jup}}$, $P_{\mathrm{orb}}$ = 1.6 d (a = 0.03 au);
    \item $M_{\mathrm{pla}}$ = 2 $M_{\mathrm{Jup}}$, $P_{\mathrm{orb}}$ = 12.5 d (a = 0.12 au);
\end{itemize}
Using M11A's inclination value of $i$ = 70$^{\circ}$, we get: 
\begin{itemize}
    \item $M_\mathrm{{pla}}$ = 1 $M_{\mathrm{Jup}}$, $P_{\mathrm{orb}}$ = 5 d (a = 0.062 au);
    \item $M_\mathrm{{pla}}$ = 2 $M_{\mathrm{Jup}}$, $P_{\mathrm{orb}}$ = 35 d (a = 0.23 au);
\end{itemize}
In a case of a star of 1 $M_{\mathrm{\odot}}$ with a transiting exoplanet, we can hope to detect a 1 $M_{\mathrm{jup}}$ orbiting at up to 10 days, using typical DI non-stabilised observations. This is, of course, given similar conditions in terms of data quality and quantity, observing constraints, and stellar variability level. As more numerous and precise RVs should be easily obtainable, it is fair to expect better results and identify HJs around very young Solar-type stars. \par

\subsubsection{Dependence of planet detection to various parameters}
\label{subsec:planetdetection}
In terms of semi-amplitude, our detection threshold of around 100\,\mos\, corresponds to half of the stellar activity rms and a quarter of its semi-amplitude ($\approx 400$ to $500$\,\mos\, looking at the maximum of the datapoints, or 357$\pm\,100$\,\mos according the the GP applied to Dataset \#5). Given the scarcity of planets orbiting very young stars discovered solely using RV, comparisons with the literature are limited. When excluding searches in the small activity regimes (i.e. $\mathrm{rms_{activity}}<50$\,\mos) only two planets provide a direct comparison, V830 Tau b \citep{Donati2017} and TAP 26 b \citep{Yu2017b}. \par 

TAP 26 b is thought to have a semi-amplitude of 160\,\mos, or $\frac{1}{8}$ to $\frac{1}{12}$ of the stellar variability semi-amplitude and V830 Tau b ($K\approx60$\,\mos) up to $\frac{1}{20}$. They both exhibit activity levels with a semi-amplitude of $\approx1200$\,\mos\,. We believe the difference in performance (detection threshold of $\approx \frac{1}{4}$ of the activity level for this work) can be explained by the fact that (i) both  \cite{Yu2017b} and \cite{Donati2017} had more data ($\approx$ 30 and 60 epochs vs 23 for us), (ii) \cite{Donati2017} had slightly better uncertainties on the RVs ($\sigma_{\mathrm{RV}}\approx50$\,\mos vs 75 for both \cite{Yu2017b} and this study) and most importantly, (iii) both had a longer baseline for the observations: 100 and 35 stellar rotation cycles vs 3 for us. We also used less constraining priors as previous knowledge was not available. \par
To ensure it was not our method implementation that hindered our capacity to find smaller signatures from our datasets, we ran our code on both \cite{Donati2016}'s and \cite{Yu2017a}'s data and were able to retrieve the published periodic signatures. We note that with no access to previous knowledge, our priors distributions were less restrictive (i.e. non Gaussian and/or broader for the concerned parameters), decreasing the evidence and yielding slightly more conservative results. Our limitations can be seen as an upper boundary and that data of better quality and quantity would be able to detect smaller planets. \par
We find that detecting planets with orbital periods conflicting (i.e. within 0.1 d of) either $P_{\mathrm{eq}}$ or its harmonics was unreliable, as illustrated in Figure~\ref{fig:result_DI} and~\ref{fig:result_GP}. Although longer period planets did not seem to be harder to detect, we noticed a significant loss of precision in the orbital period retrieved once we reach periods larger than 40 to 50 per cent of the observing time frame (see MAP values on Figure~\ref{fig:result_GP}). This is expected and good practice to sample a few orbital period at least to get reasonable constraint. A good example of a similar study can be found in \cite{Klein2020}, where the authors required 35/50 datapoints spread over 3 months ($\approx$ 15 orbital periods) to reliably detect 5/10~\mos planets behind stellar activity (about 2/3 times the planetary signature level). Finally, as we saw for datasets \#32, \#36 (see Appendix~\ref{sec:appendixB} for the complete analysis of these 2 datasets) and to a lesser extent for \#2 and \#15,  spurious periodicity signatures can appear with no relation to harmonics and no obvious relation with the window function, as suggested by \cite{Nava2019}. \par
Regarding orbital phase, the significant difference in peak height between datasets \#13 (periodogram peak power = 0.3, $\Phi$ = 0.4769) and \#15 (periodogram peak power = 0.5, $\Phi$ = 0.9183) given their identical period and comparable semi-amplitude suggests that phase impacts the detection capabilities. It is not surprising that particular phases would have an impact on the periodogram as the irregular sampling can yielded different phase coverage. That being said, we did not observe a general trend with phase across all datasets. \par
Finally, data obviously plays a huge role in the detection capabilities with crucial aspects being quality, quantity and sampling. To better characterise the activity, i.e. improve the hyperparameters of the GP, it is important to optimise the sampling (spanning multiple stellar orbits with as dense and as regular sampling as possible). Another successful strategy is to apply a GP to simultaneous photometric data, or at least not too far apart so that there is not too much evolution of the stellar surface features. We also tested (see Appendix~\ref{sec:appendixD}) the improvement brought by the knowledge of the period of the orbiting planet (i.e. characterising a known transiting planet).

\section{Conclusions}
\label{sec:Conclusions}
In this paper, we assessed our capabilities to detect exoplanets behind real stellar activity signatures. We used a previously published set of observations gathered with a non-stabilised spectrograph of the young, active G star HD 141943 in which we injected simulated planets. We then utilised two distinct strategies, Doppler Imaging and a Gaussian Process regression to filter out the stellar activity variability, aiming to recover these injected planets. \par
Our dedicated treatment of stellar activity allowed significant improvement in the detection capabilities compared to J14, a planet search study done on the same star with no dedicated activity mitigation. As previously shown by e.g. \cite{Yu2017a}, these strategies are amongst the best tools we have to deal with large activity signatures. Although now widely accepted in the community, we further confirmed that dedicated treatment of the activity is crucial, and showed that we can detect short-orbit gas giants even in non-optimally sampled datasets exhibiting 50-100\,\mos RV precision. \par
We tested two alternatives to recover the RVs from the LSD line profiles (generalised normal distribution fit and first order moment),  which yielded slightly different RV time series but more importantly different uncertainties. We favoured the FOM approach but other methods such as broadened profiles could be explored.\par
With a low number of epochs acquired with a non-stabilised spectrograph, the combined use of both GP and ZDI methods enabled us to set a planet detection threshold of around 100\,\mos\, or $\approx \frac{1}{4}$ of the activity level. Injected planets under this threshold were either non-detections or would require extra observations to confirm. The limitations we faced give a good idea of the upper limit we can hope to achieve for such systems in similar conditions. \par 
Although Doppler Imaging shows less reliability than the GP, it allows us to strengthen the confidence of a finding. This lack of reliability could be explained by the fact that DI does not take into account surface variability due to the active regions appearance/disappearance. These can evolve quickly, as shown for another G type star in \cite{Petit2004}. We suggest that claiming planets around active star should be done with a dedicated treatment for stellar variability, and preferentially using a Bayesian framework for robustness and to allow a quantification of the evidence of the presence of an orbiting planet.\par
We attempted to identify some factors that could improve the likelihood to find exoplanets orbiting young stars. Larger and more precise datasets is an obvious one. Efficient sampling is also crucial, where dense sampling of the stellar rotation to better constrain the activity should be combined with coverage of multiple planetary orbits. \par
Orbital periods close to $P_{\mathrm{eq}}$ and its harmonics pose serious difficulties, and often leads to non-detections. In our case, it also appears to be a good rule of thumb to sample at least 2, even 3 orbital periods to constrain $P_{\mathrm{orb}}$ with sufficient precision. \par
Some datasets (\#2, \#15, \#32 and \#36) exhibited significant spurious peaks of mysterious origin that compete with the true planetary period, emphasising the difficulty of RV only searches. \par
Detecting young exoplanets that do not transit is difficult but essential if we want to expand the sample of massive short period exoplanets orbiting very young stars and progress toward settling the long lasting debate over their origin. This work demonstrates that we can realistically identify potential candidates for follow-up observations and even detect short-orbit gas giants planets in non-optimised datasets exhibiting large activity variability. 

\section{Future work}
\label{sec:futurework}
As follow-up of this work, and to improve precision, producing mean line profiles with either classical approaches (i.e. CCF, shift and fit) or more recent proposals (\citealt{Rajpaul2020} or \citealt{Cameron2020}), rather than with LSD for the GP analysis could be explored. Indeed, the strength of LSD is the increase in S/N it provides, at the cost of a poorer estimations of the uncertainties (usually overestimated). It is more relevant to have a better estimation of the uncertainties when it comes to the RV used for the GP rather than a boosted S/N (required for Doppler Imaging). \par
Now having a better grasp on the capabilities of these activity mitigation strategies, it is possible to study real data of young solar type stars. Many projects such as the BCool \citep{Marsden2014} or the TOUPIES \footnote{\url{https://ipag.osug.fr/Anr_Toupies//}} (\citealt{Folsom2016,Folsom2018}) surveys, aimed at characterising star using DI and ZDI, would be good starting points. \par 
Among the overwhelming number of targets observed by the TESS mission \citet{Ricker2015}, many are young Solar analogues. Careful planning of follow-up and the availability of photometry for transiting planets would drastically increase the characterisation capabilities, see Appendix~\ref{sec:appendixD}. In general, using complementary tools to diagnose the activity, such as activity indicators and photometry are strongly recommended (e.g. \citealt{Rajpaul2015,Jones2017,Oshagh2017,Kosiarek2020}).\par

\section*{Acknowledgements}
This research has been supported by an Australian Government Research Training Program Scholarship. \par 
This paper uses data acquired in 2011 with the Anglo-Australian Telescope. We would like to thank the technical staff of the Australian Astronomical Observatory for their excellent assistance during these observations. We also wish to thank the observers: A. Collier Cameron, N. Dunstone,  G. Hussain and J.C. Ramírez Vélez. \par
In the scope of this research, we used the University of Southern Queensland's (USQ) \href{https://www.usq.edu.au/hpc}{Fawkes HPC} which is co-sponsored by the Queensland Cyber Infrastructure Foundation (QCIF). \par
This research has made use of NASA's \href{https://ui.adsabs.harvard.edu/}{Astrophysics Data System} and the \href{http://simbad.u-strasbg.fr/simbad/}{SIMBAD database} operated at CDS, Strasbourg, France. \par
For this research we made use of the following Python packages: \textsc{astropy} \citep{astropy:2013}, \textsc{corner} \citep{corner}, \textsc{loguniform} (MIT licence, João Faria), \textsc{matplotlib} \citep{Hunter:2007}, \textsc{numpy} \citep{harris2020array}, \textsc{PyMultiNest} \citep{Buchner2014} and \textsc{scipy} \citep{2020SciPy}. \par
Finally, thanks to J.C. H for the insightful conversations on the science behind this research.

\section*{Data availability}
The data underlying this article will be shared on reasonable request
to the corresponding author. The full content of Table~\ref{tab:results} is available as online material.
 
%%%%%%%%%%%%%%%%%%%%%%%%%%%%%%%%%%%%%%%%%%%%%%%%%%



% \bibliographystyle{mnras}
% \bibliography{biblio} % if your bibtex file is called biblio.bib

%\begin{thebibliography}{99}
%\end{thebibliography}

%%%%%%%%%%%%%%%%% APPENDICES %%%%%%%%%%%%%%%%%%%%%
\newpage

\appendix

%% Appendix A
\section{Bayesian model selection}
\label{sec:appendixA}

To assess the significance of the presence of periodicity in our datasets, we had to compare different models: activity only versus activity + planet(s). From parameter estimation and for a given model, a MCMC (nested sampling in our case) approach gives access to the most likely set of parameters for the model along with the associated maximum likelihood value. This non normalised likelihood is, however, not comparable across models. A model with more parameters will have more flexibility and therefore the capacity to better fit the data, yielding a greater likelihood value. To circumvent this, one tries to penalise models with more parameters, following Ockham's razor (law of parsimony). The notoriously hard to compute marginal likelihood or evidence (or posterior distribution normalisation constant) naturally applies Ockham's razor and acts to penalise models with higher number of parameters. \par
We encourage the reader to refer to \citet{Robert2009} for a thorough description of this Bayesian framework, based on the work of \citet{Jeffreys1961}. We'll here give a quick overview. \par 
First let's define the quantities required for this Bayesian framework. With $\mathbf{y}$ an array containing the data points, $\boldsymbol{\theta}$ the set of parameters composing the model and $\mathcal{M}_i$ the i$\mathrm{^{th}}$ model, we define:
\begin{itemize}
    \item $\mathrm{p}(\boldsymbol{\theta}|\mathbf{y},\mathcal{M}_i)$ or $\mathcal{P}(\boldsymbol{\theta})$: Posterior distribution of the parameters.
    \item $\mathrm{p}(\mathbf{y}|\boldsymbol{\theta},\mathcal{M}_i)$ or $\mathcal{L}(\boldsymbol{\theta})$: Likelihood.
    \item $\mathrm{p}(\boldsymbol{\theta}|\mathcal{M}_i)$ or $\pi(\boldsymbol{\theta})$: Prior probability of the parameters.
    \item $\mathrm{p}(\mathbf{y}|\mathcal{M}_i)$ or $\mathcal{Z}$: Evidence / marginal likelihood / probability of data given the model / posterior distribution normalisation constant.
    \item $\mathrm{p}(\mathcal{M}_i)$ or $\pi(\mathcal{M}_i)$: Model's prior.
    \end{itemize}

The first four are linked by Bayes' theorem:

\begin{equation}
 \mathcal{P}(\boldsymbol{\theta}) = \frac{\pi(\boldsymbol{\theta}) \times \mathcal{L}(\boldsymbol{\theta})}{\mathcal{Z}}
\end{equation}

To compare models, we look at the Bayes' Factor (BF), defined for two models $\mathcal{M}_0$ and $\mathcal{M}_1$ as:
\begin{equation}
\mathrm{BF} = \frac{\mathcal{Z}_1}{\mathcal{Z}_0}\frac{\pi(\mathcal{M}_1)}{\pi(\mathcal{M}_0)}
\label{eq:8.5}
\end{equation}

In cases where nothing a priori favours one model over the other (i.e $\frac{\pi(\mathcal{M}_1)}{\pi(\mathcal{M}_0)}$ = 1), we are left with the ratio of the marginal likelihoods. To obtain the marginal likelihoods, we need to marginalise (i.e. integrate) over all parameters:

\begin{equation}
\mathcal{Z} = \int 
\mathcal{L}(\boldsymbol{\theta}) \pi(\boldsymbol{\theta}) \mathrm{d}^N \boldsymbol{\theta}
\label{eq:9}
\end{equation}

with N the number of parameters. For models with a large number of parameters, accurate computation of $\mathcal{Z}$ requires integrating over many dimensions (N) and is therefore often intractable. It quickly becomes too computationally expensive and needs to be approximated. Various approaches are used in the literature, see \citealp{Nelson_2020} for the most extensive attempt to compare these methods in the context of exoplanet searches. \par 
Once in possession of the evidence for each model we can compute the BF and assess whether the data favours model 1 over model 0. It is common practice to work with the \emph{natural} logarithm of the evidence, we then have:
\begin{equation}
\mathrm{BF} = \exp(\ln\mathcal{Z}_1 - \ln\mathcal{Z}_0) = \exp(\Delta\ln\mathcal{Z})
\label{eq:10}
\end{equation}
The different level of confidence is then interpreted from the Bayes Factor, following \citet{Jeffreys1961}, as summarised in Table~\ref{tab:Bayes factor}. We emphasise that this level of evidence is assessing the significance of the model given the data, and does not take into account how accurately the data reflects what we wish to observe nor the plausibility of the model on its own (although that can be added as $\pi(\mathcal{M}_i)$ in Equation~\ref{eq:8.5}).

\subsection{Notes on prior probabilities}
\label{subsec:priorsappendix}
Prior probabilities are at the core of Bayesian inference, and express knowledge previously acquired on a particular aspect of the problem, i.e. a parameter of the model, or on the model itself. One must be mindful of the choice for these priors. Some are `uninformative' (Uniform, Jeffrey's or Modified Jeffrey's priors, see \cite{Gregory2007} for more details on the last two), meaning that they do not contain a priori information (they do to some extent as a uniform prior has limits, but these are rather physical than inferred from previous analyses). On the other hand, so called `informative' priors, typically a Gaussian prior, can strongly constrain the parameter space to be explored. This results in a boosted marginal likelihood compared with the use of an `uninformative prior'. \par
Therefore, one has to be extremely cautious when using `informative' priors. The previous knowledge yielding that prior has to be statistically robust, to not mislead the analysis by boosting the evidence.

\subsection{Note on evidence vs. planet detection}
Accurate estimation of the evidence grants a statistically robust framework to measure the significance of \emph{one model relative to others}. We should keep in mind that models could be wrong and that drawn conclusions about physical phenomena are only justified as long as these models and their underlying assumptions are reasonable. \par
Here, our model assesses the likelihood of the presence of a periodic signature in a dataset. It does not, however, inform us about the origin of such signatures. Assumptions on the nature of the underlying physical phenomenon or the accuracy of the data collection and reduction are required to go from `a periodic signal in the data' to `a planet orbiting the observed star'. For example, an previous exoplanet detection claim was later attributed to the window function by \citet{Rajpaul2016}. It has been the case again very recently for V830 Tau b \citep{Damasso2020}, where the signature found by \cite{Donati2016,Donati2017} could not be found in a new dataset. Whether this means the planet does  or not exist is a rather challenging question, but it once more highlights the difficulty of RV only searches.\par 

\begin{table}
	\centering
	\caption{Values of the Bayes Factor, i.e. the ratio of marginal likelihood between the single planet model and the 0-planet (stellar activity only) model. Middle and left columns are the associated probability and level of confidence.}
	\label{tab:Bayesproba}
    \begin{tabular}{ccc}
    \hline
    Bayes factor & Probability & Level of confidence \\
    \hline
     < 1 & < 0.5 & None \\
     < 3 & < 0.75 & Not worth more than a bare mention \\
     < 10 & < 0.909 & Substantial \\
     < 30 & < 0.967 & Strong  \\
     < 100 & < 0.99 & Very strong \\
     > 100 & > 0.99 & Decisive\\
     
    \label{tab:Bayes factor}
    \end{tabular}
\end{table}

\newpage 
%% Appendix B

\section{Spurious periodogram peaks}
\label{sec:appendixB}

\begin{figure*}
    \centering
    \includegraphics[]{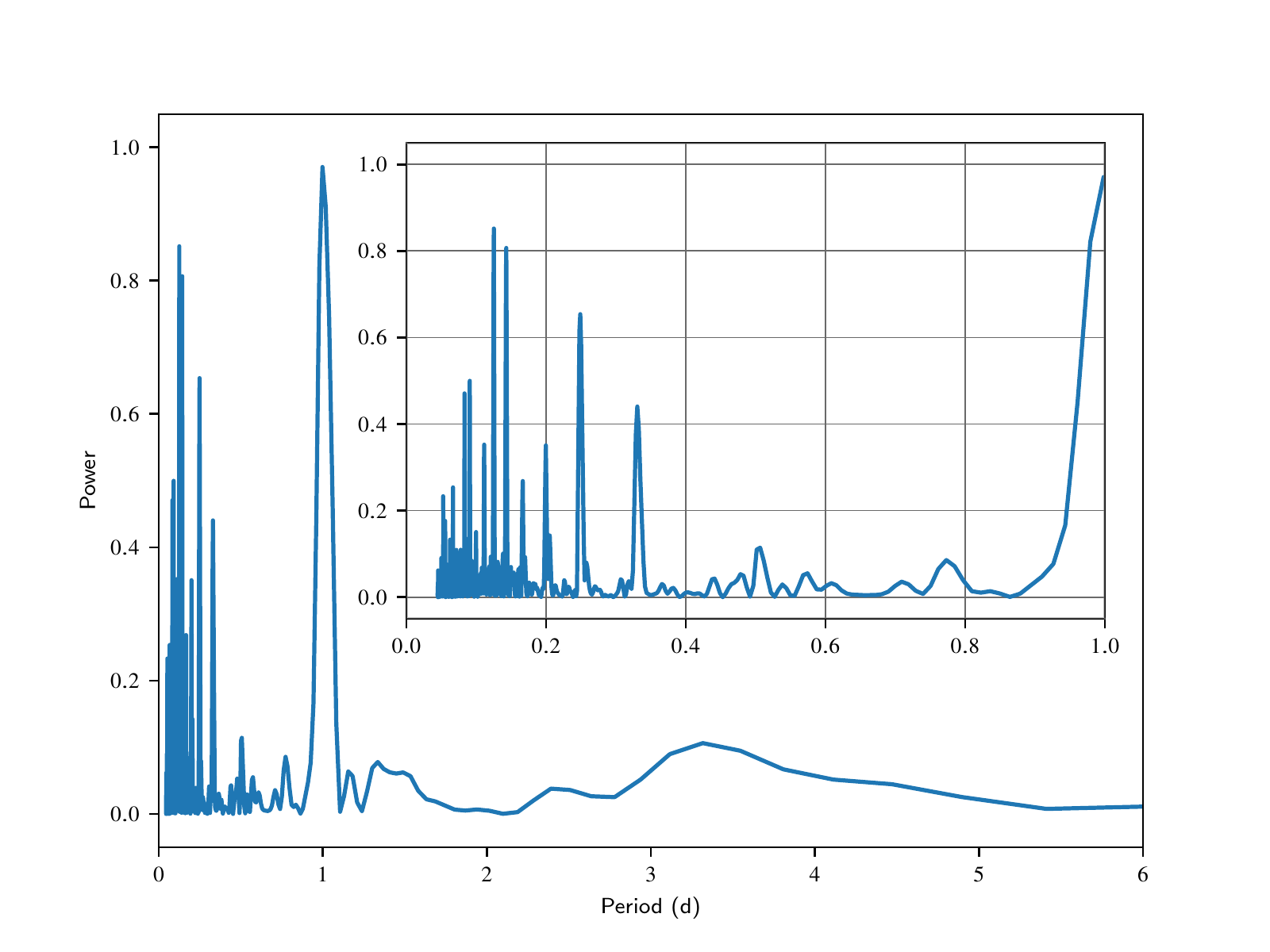}
    \caption{Window function of the observed RVs. The horizontal axis shows the period in days and the vertical axis the periodogram power. The upper-right zoomed window shows extra details for periods < 1 day.}
    \label{fig:window}
\end{figure*}

\begin{figure*}
    \includegraphics[width=0.75\textwidth]{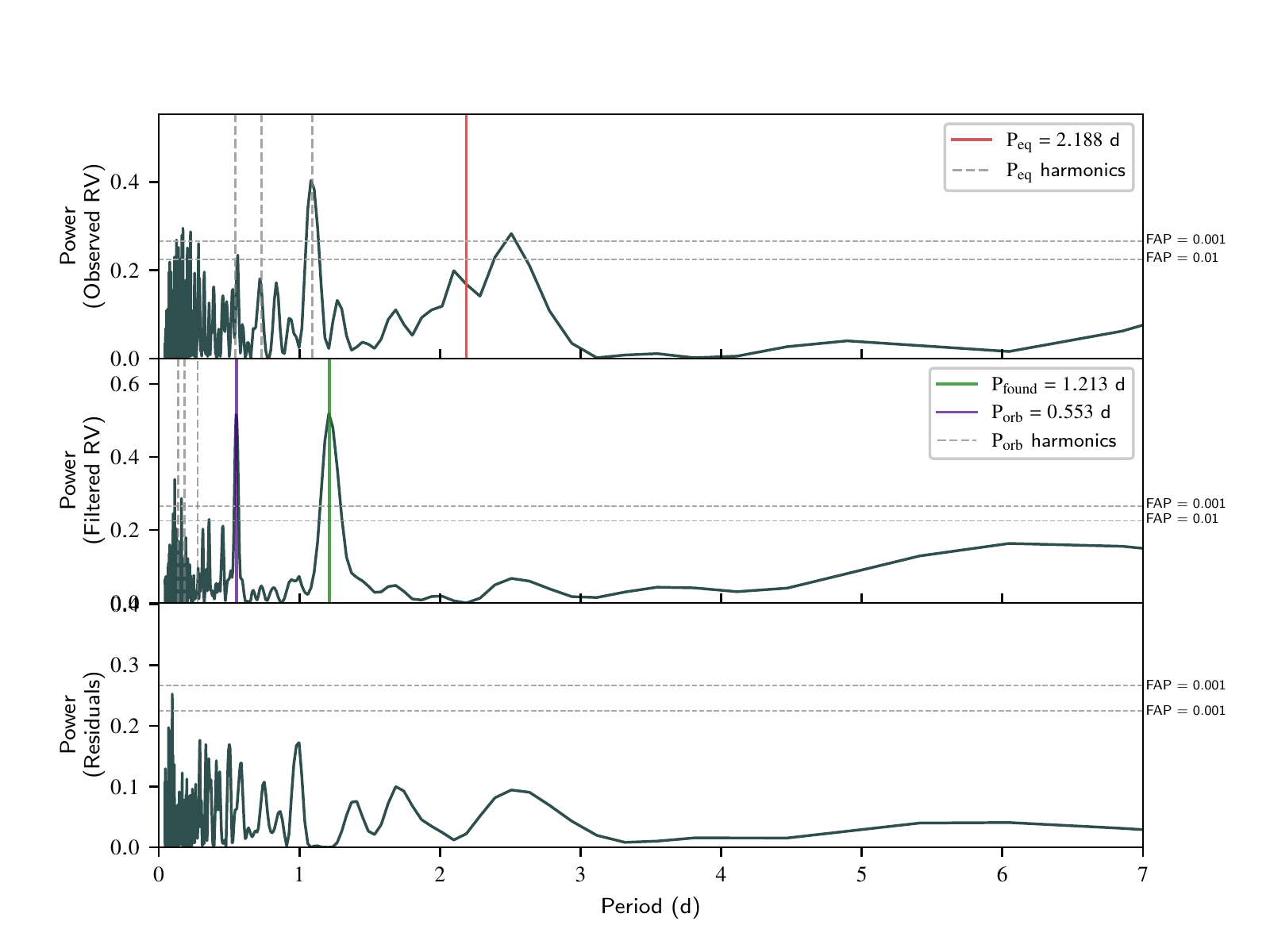}
    \includegraphics[width=0.75\textwidth]{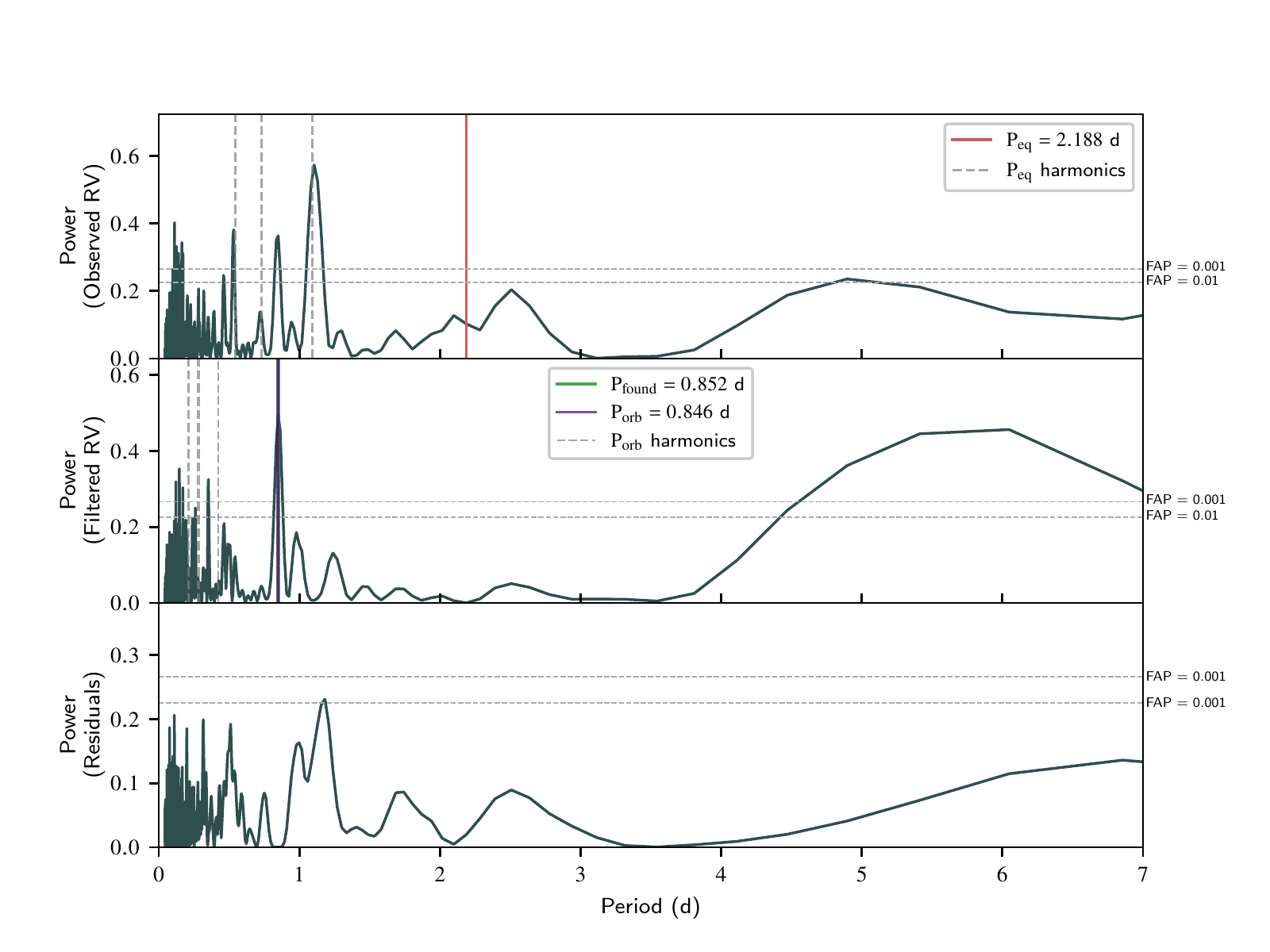}
    \caption{Periodograms of datasets \#32 (top) and \#36 (bottom). Each plot has 3 subplots showing the periodograms of the raw observed data (top), DI filtered (middle), i.e observed - synthetic RVs and residuals (bottom), i.e. filtered - identified periodic signature. Vertical lines on the top subplots mark the stellar rotation period and its harmonics. In the middle subplots, we showed the recovered periodicity (green vertical line) and the true injected period (purple vertical line), along with its harmonics (grey dashed vertical lines).}
    \label{fig:periodograms32_36}
\end{figure*}

For the 2 following datasets:
\begin{itemize}
    \item \#32: $K$ = 149.72\,\mos, $P_{\mathrm{orb}}$ = 0.5526 d and $\Phi$ = 0.5701
    \item \#36: $K$ = 142.36\,\mos, $P_{\mathrm{orb}}$ = 0.8463 d and $\Phi$ = 0.6225
\end{itemize}
the periodogram resulting from the DI analysis showed, in each case and after the activity filtering, the apparition of a spurious peak at periods seemingly unrelated to either the stellar rotation harmonics, the orbital period harmonics or a peak in the window function (see Figure~\ref{fig:window}). As shown in Figure~\ref{fig:periodograms32_36}, $P_{\mathrm{spurious32}} \approx$ 1.22 d and $P_{\mathrm{spurious36}} \approx$ 5.7 d. After removing the identified highest peak from the filtered data (matching with the true period in the case of \#36 but not for \#32) the competing peak was also filtered out, suggesting an effect of the uneven sampling of the observations. \par
\begin{figure*}
    \includegraphics[width=\textwidth]{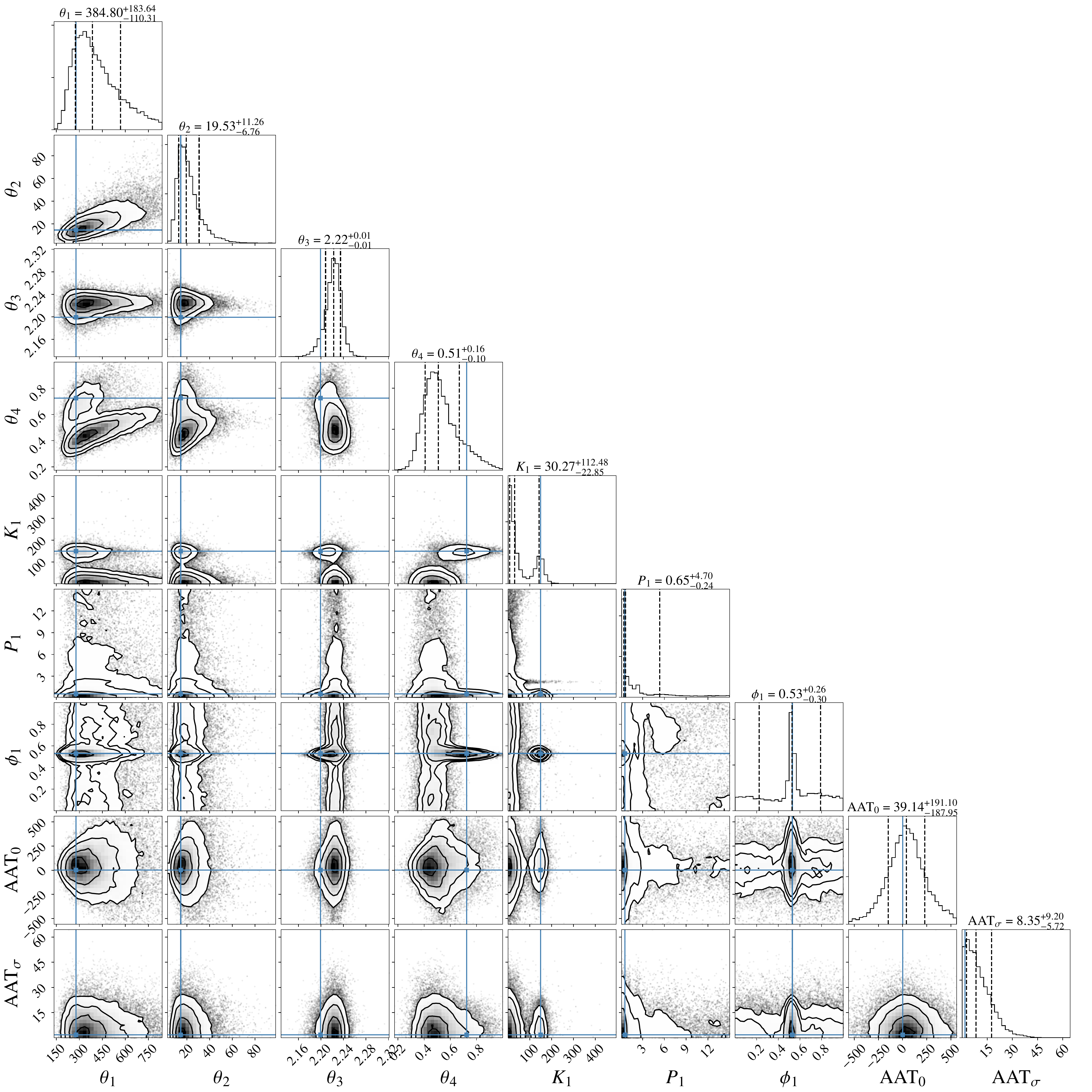}
    \caption{Posterior distribution of the parameters from the} GP analysis of dataset \#32. Prior weighted posterior samples drawn from the \textsc{pymultinest} analysis. Blue lines show the most likely parameters. Dashed vertical lines are 0.16, 0.5 and 0.84 quantiles. Contours are 1, 2 and 3 sigma levels (representing respectively 39, 86 and 99 per cent for a 2D distribution).
    \label{fig:corner32}
\end{figure*}
\begin{figure*}
    \includegraphics[width=\textwidth]{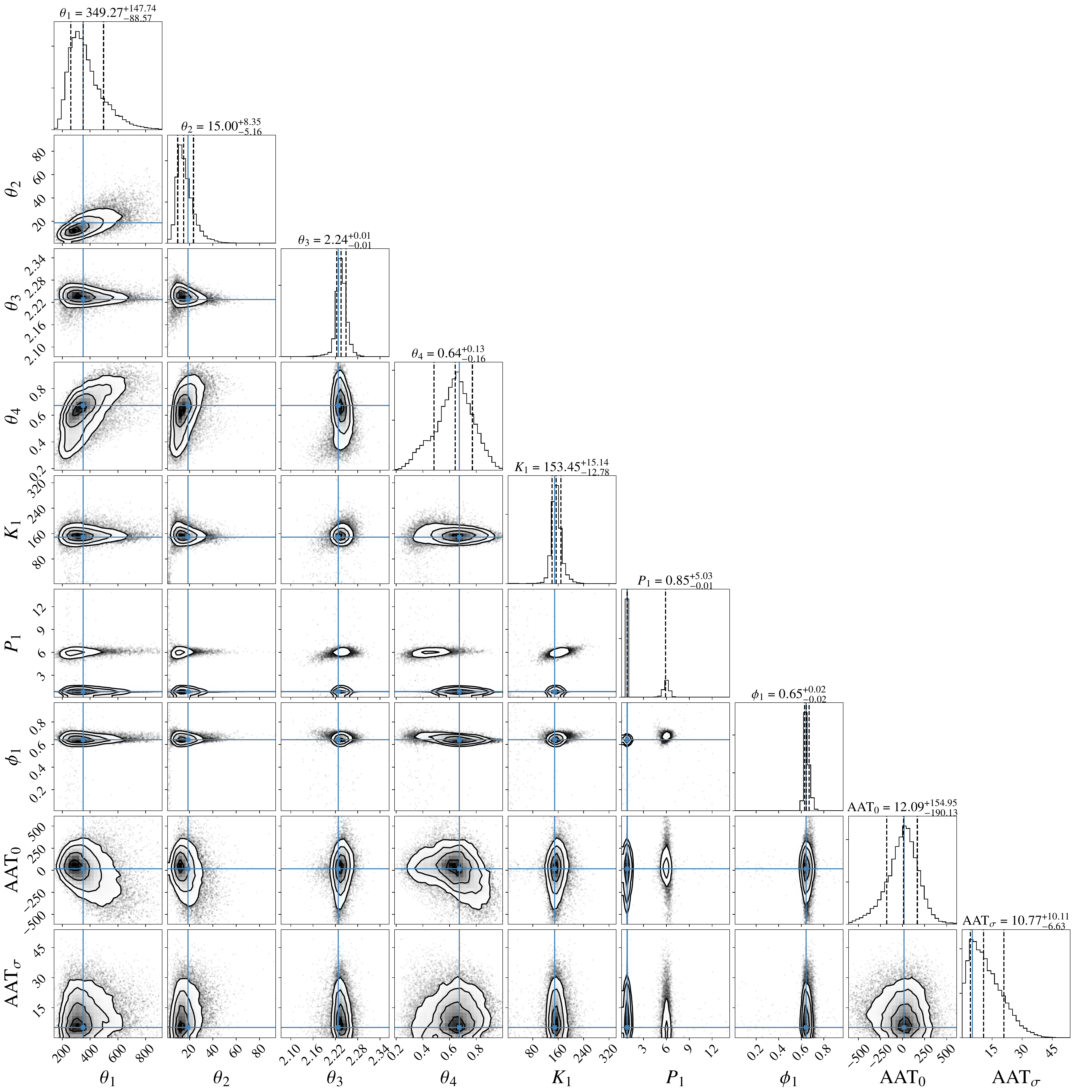}
    \caption{Posterior distribution of the GP analysis of dataset \#36. Prior weighted posterior samples drawn from the \textsc{pymultinest} analysis. Blue lines show the most likely parameters. Dashed vertical lines are 0.16, 0.5 and 0.84 quantiles. Contours are 1, 2 and 3 sigma levels (representing respectively 39, 86 and 99 per cent for a 2D distribution).}
    \label{fig:corner36}
\end{figure*}
\begin{figure*}
    \includegraphics[width=0.8\textwidth]{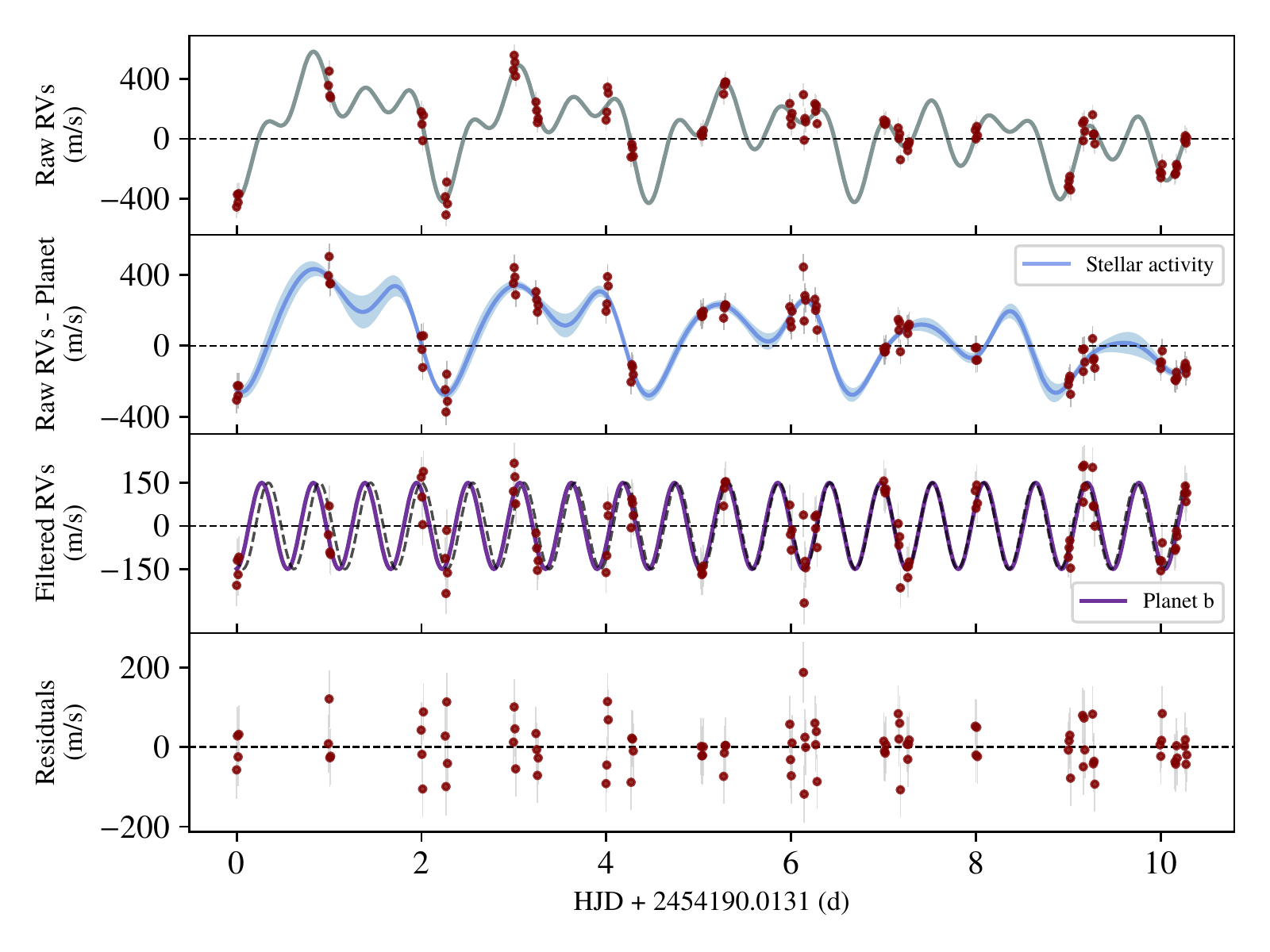}
    \includegraphics[width=0.8\textwidth]{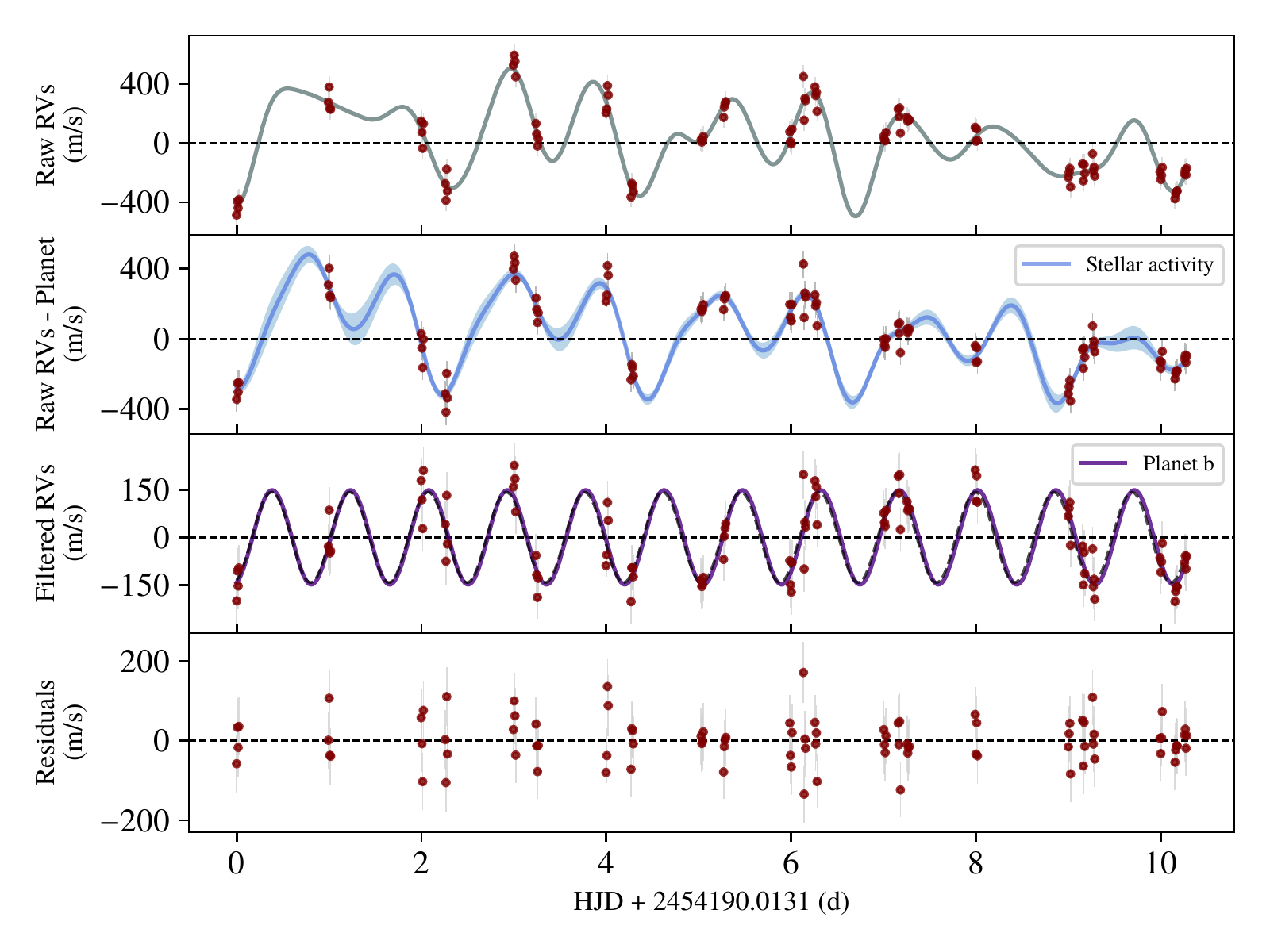}
    \caption{Resulting fits for the GP analysis of dataset \#32 (top) and \#36 (bottom). Top panel is the total model (grey line) and the observed raw RVs (red points). The second panel is the GP fit with the uncertainty shown by the shadowed area (analytically computed predictive standard deviation of the GP), red points are (raw RVs) - (recovered planet). The third panel is the recovered planet (purple line) on top of the true injected planet (dashed curve). Red points are  (raw RVs) - (GP fit). The bottom panel are the residuals, i.e. (raw RVs) - (GP model) - (recovered planet).}
    \label{fig:result_32_36}
\end{figure*}

Figures~\ref{fig:corner32},~\ref{fig:corner36} and~\ref{fig:result_32_36} show the corner plot of the posterior distribution and the resulting fit to the data for these 2 datasets following the GP analysis. \par 
For dataset \#32 (Figure~\ref{fig:corner32} and top plot in Figure~\ref{fig:result_32_36}), the GP does not seem to be fooled and clearly identifies the right period. However, the evidence is surprisingly low for such a large planet (BF = 1.67 or p = 0.62 in favour of the single model), yielding a non-detection. We strongly suspect that the ambiguity in both methods is due to the orbital period of the injected planet being very close to the third harmonic of the stellar period ($\frac{P_{\mathrm{eq}}}{3}$ = 0.547 d and $P_{\mathrm{orb}}$ = 0.553 d). \par 
Regarding dataset \#36 (Figure~\ref{fig:corner36} and bottom plot in Figure~\ref{fig:result_32_36}), once again the correct period is identified by the GP. We also note a slight appearance of the conflicting period around 6 days on the posterior distribution (Figure~\ref{fig:corner36}). This time however, the evidence is strongly in favour of a detection with a Bayes Factor of 175.9 (p = 0.994). \par
Although it is not clear where the spurious peak arise from, one of these two cases could be settled by the GP. 

%% replace with corner plots

\clearpage
%% Appendix C
\section{Impact of different DI solutions on Dataset \#22}
\label{sec:appendixC}

As we saw in Section~\ref{subsec:refinedstellarparam}, the stellar parameters found via Doppler imaging are slightly different from M11A/B. Because the DI filtering uses the synthetic generated line profiles that depend on the stellar parameters, we tested the influence of three different DI solutions on our planet detection capabilities. We re-analysed dataset 22, containing a injected planet with $K$ = 117.28\,\mos, $P_{\mathrm{orb}}$ = 3.1546 d and $\Phi$ = 0.7077. \par
As described in Section~\ref{subsec:refinedstellarparam}, we derived the alternate Doppler Imaging solutions by fitting (i) only using dark features and (ii) using both dark and bright features but forcing the inclination parameter to 70$^{\circ}$ (matching M11A/B's value). The optimum parameters are displayed in Table~\ref{tab:bestparam22}. \par
\begin{figure*}
    \centering
    \includegraphics[width=\textwidth]{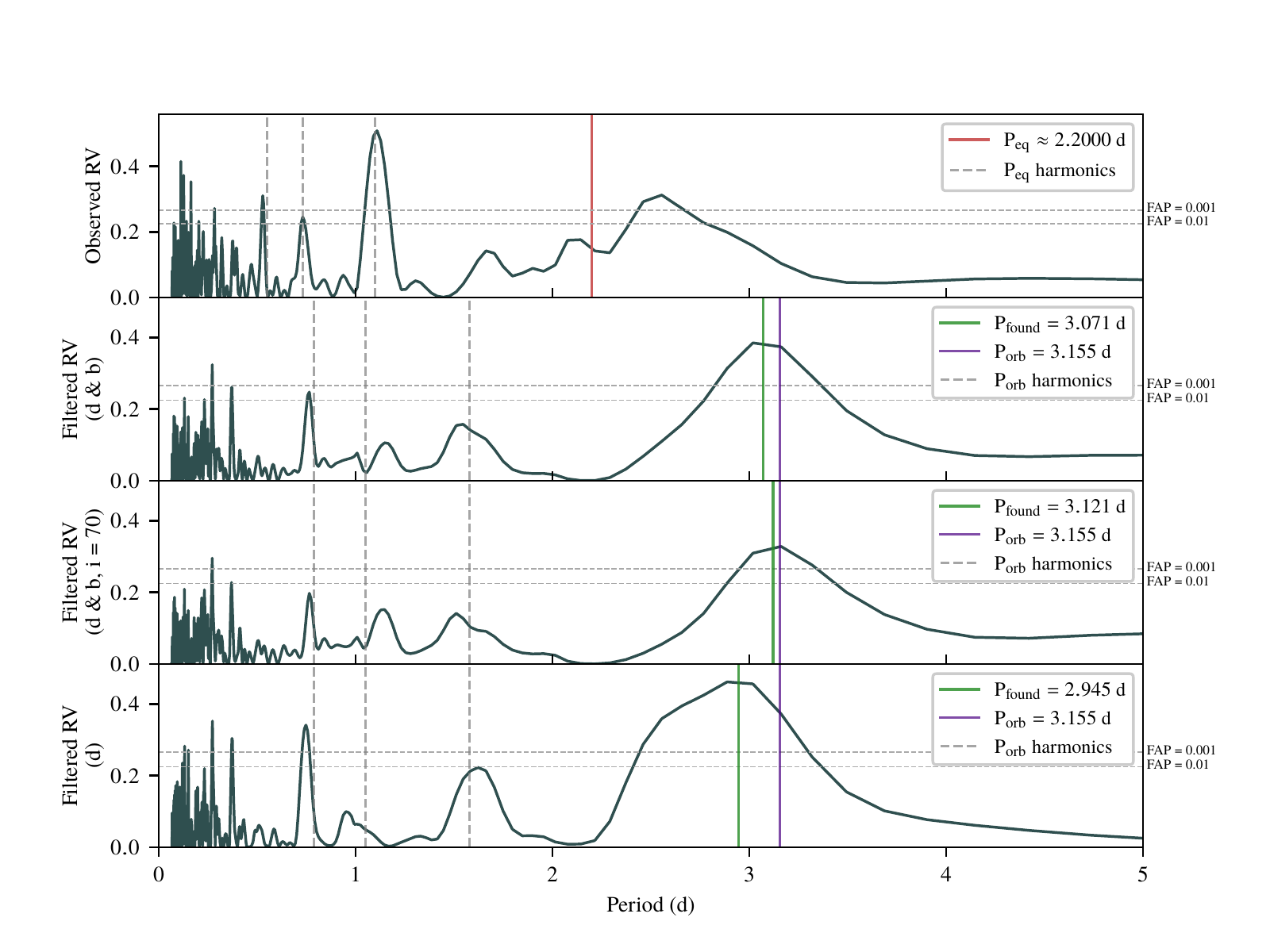}
    \caption{Top: periodogram of the raw data from Dataset \#22. Second to fourth plot: Comparison of the periodograms for dataset \#22 obtained from dark and bright spot DI analysis (2$\mathrm{^{nd}}$, d \& b), from dark and bright spot DI analysis with imposed inclination (3$\mathrm{^{rd}}$, d \& b, i = 70) and from dark spots only DI analysis (4$\mathrm{^{th}}$, d).}
    \label{fig:per_22}
\end{figure*}

 
Regarding the maps, results are similar to Dataset \#5 (no planet), with mainly a difference in contrast. The periodograms of the filtered RVs (synthetic RVs - observed RVs) for each of the three cases are displayed in Figure~\ref{fig:per_22}. In all cases, the features in the periodogram are quite similar. The analysis with dark spots only is the most different. It appears, in that case, that the original peak due to activity was not completely filtered and blends with the peak of the injected period after the filtering process. This acts to slightly hinder the accuracy of the period retrieved. \par
The fixed inclination case performs slightly better than the dark and bright case, as the retrieved period is 0.03 d away from the injected period for the former versus 0.08 d for the latter. However, the shape of the peak is very similar and we cannot conclude whether the better result is in fact due to a better solution for the DI or not.

\begin{table}
	\centering
	\caption{Test of different DI solutions fitting for dark and bright spots (top row), dark and bright features with the inclination parameter forced to 70$^{\circ}$ (middle row) and only dark features (bottom row).}
	\label{tab:bestparam22}
    \begin{tabular}{lllll}
    \hline
    Fitting for & $v\sin i$ & $P_{\mathrm{eq}}$ & $\mathrm{d}\Omega$ & \radd \\
    \hline
    Dark and bright & 35.54 & 2.2003 & 0.1231 & 47 \\
    Dark and bright (fixed $i$) & 35.58 & 2.1982 & 0.1138 & 70 (fixed) \\
    Dark & 35.38 & 2.2115 & 0.0441 & 44 \\
    \hline
    \end{tabular}
\end{table}

\clearpage 

%% Appendix C
\section{Simulating a transiting planet}
\label{sec:appendixD}
In order to showcase what access to a photometrically detected planet (i.e. the case of a transiting planet) would add to the retrieval capabilities, we investigated a dataset allowing for additional prior constraints. For this we chose dataset \#29, which exhibits the smallest semi-amplitude among the datasets that are not close to either $P
_{\mathrm{eq}}$ or its harmonics. It has a semi-amplitude of 47\,\mos, well below our detection threshold. In Table~\ref{tab:GP29}, we compared the evidences (derived from the GP analysis) for 3 different cases. \par 
The first case (first line of Table~\ref{tab:GP29}) uses the same priors as we did throughout this article, simulating access solely to RV data. The second case (second line of Table~\ref{tab:GP29}) simulates the availability of transiting data on the planet. We set Gaussian priors for $P_{\mathrm{orb}}$ and $\Phi$, centred on the true injected value and with standard deviation of respectively 0.0001 d and 0.01. For the third case (third line of Table~\ref{tab:GP29}), we fixed the value for $P_{\mathrm{orb}}$ and $\Phi$ to their true values (simulating the best transit value). \par
For each cases, the Bayes Factor and probability favouring the single planet model over the activity only model can be found in the last two columns of Table~\ref{tab:GP29}. For the first case, the BF is extremely low (0.4, p = 0.27) and leads to a non-detection (see grey cross labelled `29' on Figure~\ref{fig:result_GP}). For cases 2 and 3 however (lines 2 and 3 of Table~\ref{tab:GP29}), their BF are comparable and around 9, meaning a 0.9 probability in favour of the 1-planet model and therefore a strong evidence for the presence of a planet. \par 
This is expected as our constrained priors act to boost the evidence. It also 
shows how an inappropriate choice of priors can influence the evidence and bias the claim of a finding, following our discussion in \ref{subsec:priorsappendix}. We see once more the difficulty of RV only searches.

\begin{table*}
	\centering
	\caption{GP analysis of dataset \#29 with 3 different set of priors. The first column shows the real case configuration leading to the according choice of priors. The second column is the list of parameters with a Gaussian prior. The third column is the list of fixed parameters, i.e. not taking part in the parameter space search process. The last 2 columns are the Bayes Factor and the probability favouring the single planet model over the activity only model resulting from the corresponding analysis. To constrain or fix $P_{\mathrm{orb}}$ and $\Phi$, we used the true injected values as they would be available if transits would have been identified.}
	\label{tab:GP29}
    \begin{tabular}{lllll}
    \hline
    Analysis & Constrained (i.e. Gaussian prior) & Fixed (not fitted) & BF & Probability \\
    \hline
    RV only & $\theta_3$ & None & 0.4 & 0.27 \\
    Transit &  $\theta_3$, $P_{\mathrm{orb}}$, $\Phi$ & None & 9.9 & 0.91\\
    Transit (fixed) & $\theta_3$ & $P_{\mathrm{orb}}$, $\Phi$ & 8.2 & 0.89 \\
    \hline
    \end{tabular}
\end{table*} 

\label{lastpage}
\end{document}